\documentclass[12pt,a4paper]{article}
\pdfoutput=1
\usepackage{color}
\usepackage{amssymb,amsmath,bbold,MnSymbol}
\usepackage{epsf}
\usepackage{epsfig}
\usepackage{relsize}
\usepackage[dvipsnames]{xcolor}
\usepackage[linktoc=page,bookmarks=false,colorlinks=false,linkbordercolor=RoyalBlue,citebordercolor=ForestGreen,urlbordercolor=CornflowerBlue]{hyperref}
\usepackage{latexsym,mathrsfs,dsfont}
\usepackage[normalem]{ulem} 
\usepackage[compress]{cite}
\usepackage{graphicx}
\usepackage{url}
\usepackage{booktabs}
\usepackage{float}
\usepackage{multirow}
\usepackage{changepage}
\usepackage[hypcap]{caption, subcaption}
\usepackage{adjustbox}

\usepackage[titles]{tocloft}

\usepackage{tabularx,colortbl}

\usepackage{fancyhdr}
\advance \headheight by 3.0truept       

\allowdisplaybreaks[1]

\definecolor{red}{cmyk}{0,1,1,0.4}
\definecolor{darkgreen}{rgb}{0.0,0.6,0.0}
\definecolor{cDarkGrey}{RGB}{91,91,91}
\definecolor{cGrey}{RGB}{245,243,238}
\definecolor{cBlue}{RGB}{0,110,191}
\definecolor{cLightBlue}{RGB}{214,237,252}
\definecolor{cRed}{RGB}{196,0,100}
\definecolor{cLightRed}{RGB}{254,222,237}
\definecolor{cGreen}{RGB}{0,166,80}
\definecolor{cLightGreen}{RGB}{254,222,237}
\definecolor{cOrange}{RGB}{221,74,44}
\definecolor{cLightOrange}{RGB}{255,215,210}
\definecolor{cPurple}{RGB}{93,35,125}
\definecolor{cLightPurple}{RGB}{241,230,252}
\definecolor{cYellow}{RGB}{252,191,10}
\definecolor{cISSRBlue}{RGB}{0,111,174}
\definecolor{cISSRGrey}{RGB}{167,169,172}

\newcommand{\beq}{\begin{equation}}
\newcommand{\eeq}{\end{equation}}
\newcommand{\be}{\begin{equation}}
\newcommand{\ee}{\end{equation}}
\newcommand{\bi}{\begin{itemize}}
\newcommand{\ei}{\end{itemize}}
\newcommand{\ba}{\begin{array}}
\newcommand{\ea}{\end{array}}
\newcommand{\beqa}{\begin{eqnarray}}
\newcommand{\eeqa}{\end{eqnarray}}
\newcommand{\bea}{\begin{eqnarray}}
\newcommand{\eea}{\end{eqnarray}}
\newcommand{\beqn}{\begin{eqnarray}}
\newcommand{\eeqn}{\end{eqnarray}}

\newcounter{TODO}

\def \refsec#1{Section~\ref{#1}}

\def \refapp#1{Appendix~\ref{#1}}
\def \reffig#1{Figure~\ref{#1}}

\def \reftab#1{Table~\ref{#1}}

\def \One{\text{\normalsize $\mathbb{1}$}}

\newcommand{\bm}{\boldmath}

\newcommand{\oL}[1]{\overline{#1}}

\newcommand{\MSbar}{${\overline{\text{MS}}}$}

\newcommand{\DF}{\Delta F}

\newcommand{\cL}{\mathcal{L}}

\newcommand{\cO}{\mathcal{O}}

\newcommand{\GeV}{\,\text{GeV}}

\newcommand{\epe}{\varepsilon'/\varepsilon}

\newcommand{\KKbar}{K^0-\bar K^0}

\newcommand{\alS}{\alpha_s}

\newcommand{\alE}{\alpha_\text{em}}

\newcommand{\muLow}{{\mu_\text{had}}}
\newcommand{\muEW}{{\mu_\text{ew}}}
\newcommand{\muNP}{{\Lambda}}

\newcommand{\OpL}[2][{}]{Q_{#2}^{#1}}
\newcommand{\opL}[3][{}]{[Q_{#2}^{#1}]_{#3}}

\newcommand{\adm}{\hat\gamma}
\newcommand{\DelrS}{\Delta \hat r_s}
\newcommand{\DelrE}{\Delta \hat r_\text{em}}

\newcommand{\SUthreeC}{{\mathrm{SU(3)_c}}}
\newcommand{\SUtwoL}{{\mathrm{SU(2)_L}}}
\newcommand{\UoneY}{{\mathrm{U(1)_Y}}}
\newcommand{\UoneEM}{{\mathrm{U(1)_{\rm em}}}}

\newcommand{\Nc}{{N_c}}
\newcommand{\Nf}{{N_f}}
\newcommand{\Nu}{{N_u}}
\newcommand{\Nd}{{N_d}}

\newcommand{\eps}{\epsilon}

\begin{document}

\begin{flushleft}
{\em Version of \today}
\end{flushleft}

\vspace{-14mm}
\begin{flushright}
  AJB-21-5\\
\end{flushright}

\medskip

\begin{center}
{\LARGE\bf\boldmath
  General non-leptonic $\DF=1$ WET  \\[0.3cm]
  at the NLO in QCD
}
\\[1.2cm]
{\bf
  Jason~Aebischer$^{a}$,
  Christoph~Bobeth$^{b}$,
  Andrzej~J.~Buras$^{b}$,
  Jacky Kumar$^{b}$ and
  Miko{\l}aj Misiak$^c$
}\\[0.5cm]

{\small
$^a$Department of Physics, University of California at San Diego,
    La Jolla, CA 92093, USA \\[0.2cm]
$^b$TUM Institute for Advanced Study,
    Lichtenbergstr. 2a, D-85747 Garching, Germany \\[0.2cm]
$^c$Institute of Theoretical Physics, Faculty of Physics,
    University of Warsaw, \\
    Pasteura 5, 02-093 Warsaw, Poland. \\[0.2cm]
}
\end{center}

\vskip 1.0cm

\begin{abstract}
\noindent

We reconsider the complete set of four-quark operators in the
Weak Effective Theory (WET) for non-leptonic $\DF=1$ decays that
govern $s\to d$ and $b\to d, s$ transitions in
the Standard Model (SM) and beyond, at the Next-to-Leading Order
(NLO) in QCD. We discuss cases with different numbers $\Nf$ of active flavours,
intermediate threshold corrections, as well as the issue of
transformations between operator bases beyond leading order to
facilitate the matching to high-energy completions or the Standard
Model Effective Field Theory (SMEFT) at the electroweak scale.
As a first step towards a SMEFT NLO analysis of $K\to\pi\pi$ and
non-leptonic $B$-meson decays, we calculate the relevant WET Wilson
coefficients including two-loop contributions to their renormalization
group running, and express them in terms of the Wilson coefficients in
a particular operator basis for which the one-loop matching to SMEFT
is already known.

\end{abstract}

\setcounter{page}{0}
\thispagestyle{empty}
\newpage

\setcounter{tocdepth}{2}
\setlength{\cftbeforesecskip}{0.21cm}
\tableofcontents

\newpage

%
%
%
\section{Introduction}

Non-leptonic $\DF=1$ decays mediated by down-type $b\to d, s$ and
$s\to d$ transitions offer some of the most important probes of the
quark-mixing mechanism in the Standard Model~(SM). In particular, they
play an important role in the Cabibbo-Kobayashi-Maskawa
(CKM)~\cite{Cabibbo:1963yz,Kobayashi:1973fv} matrix determination,
and enable a detailed understanding of CP violation in the quark
sector. The most prominent examples are the ``gold-plated'' decays
$B_d\to J/\psi\, K^{(*)}$ and $B_s \to J/\psi\, \phi$ that are driven
by the $b\to sc\bar{c}$ tree-level transitions. Interference of their
amplitudes with time-dependent $\DF=2$ meson mixing
amplitudes gives the most precise determination of the CKM
angles known as $\beta_d$ and $\beta_s$, respectively.  Another
example are the measurements of the CKM angle $\gamma$ in $B \to DK$
decays mediated by tree-level $b\to cud$ transitions.  Probably one of
the most sensitive observables for CP-violating effects beyond the SM
in quark-flavour physics is the ratio $\epe$ in $K\to\pi\pi$ decays
where a large hierarchy among magnitudes of the CKM elements
makes the SM contribution strongly suppressed. A recent review of
these topics and the present status of non-leptonic meson decays can
be found in Ref.~\cite{Buras:2020xsm}. See also
Refs.~\cite{Artuso:2015swg, Fleischer:2018bld, Lenz:2019lvd,
Lenz:2020awd}.

The main obstacle to fully profiting from the potential of
non-leptonic decays in tests of the SM and searches for
New Physics (NP) is our imprecise knowledge of
nonperturbative hadronic bound-state effects. In the SM,
the left-handed nature of weak interactions leads to
a specific form of the corresponding Weak
Effective Theory (WET)\footnote{
In the literature, the name LEFT (Low-energy Effective
Field Theory) is often used instead of WET.}
that allows us to develop specially tailored strategies for
elimination of uncertainties that stem from such bound-state
effects in the aforementioned non-leptonic decays. However,
contributions beyond the SM (BSM) can upset such strategies.
Firstly, they may introduce additional CP-violating phases.
Secondly, they may give rise to additional operators in
the WET, thereby increasing the number of unknown
observable-specific hadronic matrix elements of operators. The latter
are governed by strong dynamics, which requires first-principle
nonperturbative methods for their evaluation.

The available accuracy of first-principle methods in
calculations of hadronic matrix elements is strongly
process-dependent. Several new lattice results for $K\to\pi\pi$
became available over the last five years~\cite{Bai:2015nea,
Blum:2015ywa, Abbott:2020hxn}. Further insight can be gained
from large-$\Nc$ QCD calculations~\cite{Buras:2014maa,
Aebischer:2018rrz}. A good description of the $\Delta I = 1/2$
rule \cite{Abbott:2020hxn, Buras:2014maa}
and improved predictions for $\epe$ have been
obtained~\cite{Abbott:2020hxn, Aebischer:2020jto}.  Two-body decays
of $B$ mesons can be treated with QCD-factorization
methods~\cite{Keum:2000wi, Beneke:2000ry, Beneke:2001ev} that
are based on an expansion in inverse powers of the heavy-quark
mass $m_b$. The leading-power contributions depend on a few
universal hadronic matrix elements. They usually provide a good
qualitative description of decay rates. However, predictions
for CP-violating observables often suffer from unknown
subleading-power corrections, for which a proper factorization has
not yet been established. In consequence, precision analyses can
hardly be performed. Certain observables in $B$~physics can be
treated with the Heavy Quark Expansion (HQE),
as, for example, the decay-width difference
$\Delta\Gamma_{d,s}$ of neutral $B$~mesons~\cite{Artuso:2015swg,
Mannel:2020ups}. The corresponding hadronic matrix elements are
accessible to lattice methods~\cite{Davies:2019gnp}, and high
precision is expected in the future~\cite{Mannel:2020fts,
Lenz:2020oce}.  On the other hand, a rigorous description of
bound-state effects in non-leptonic charm decays is
currently not available~\cite{Khodjamirian:2017zdu}.

A necessary prerequisite is the knowledge of the WET Wilson
coefficients (effective couplings) at the relevant low-energy scale
$\muLow \lesssim 5\GeV$ at which the non-leptonic processes of interest take
place. They must be determined with the help of
Renormalization-Group (RG) improved perturbation
theory. The first necessary step is a determination of
the initial Wilson coefficients through the matching to a UV
completion at the electroweak scale $\muEW\sim 100\GeV$. Secondly,
Anomalous Dimension Matrices (ADMs) have
to be calculated. They originate from QCD and QED interactions
associated with the residual $\SUthreeC \otimes \UoneEM$ gauge
symmetry. The RG evolution of the Wilson coefficients from
$\muEW$ to $\muLow$ is governed by the ADMs. When crossing the
quark-decoupling thresholds, the corresponding corrections
need to be included in the Next-to-Leading Order (NLO) RG
evolution. This concerns, in particular, the $\Delta S = 1$
decays for which subsequent decoupling of the bottom and the charm
quarks implies working with $\Nf = 4,3$ active flavours.\footnote{
Currently most lattice calculations of $K\to\pi\pi$ matrix
elements are done with $\Nf = 3$ active quark flavours. In the
future, treating the charm quark as dynamical should
allow to raise $\muLow$ sufficiently above $1\GeV$, into a more
perturbative regime, and work with $\Nf = 4$ only.}
Eventually, in the last step, the aforementioned hadronic matrix
elements appearing in specific observables need to be evaluated with
nonperturbative methods.

Except for the latter step, calculations of the QCD and QED effects
can systematically be performed within perturbation theory, enabling
a precise calculation of the Wilson coefficients at $\muLow$.
Cancellations of intermediate renormalization-scheme dependences are
well understood. The remaining final scheme dependence gets cancelled
in predictions for observables by the corresponding scheme dependence
of hadronic matrix elements. In fact, based on the universal
character of the WET,
\footnote{As long as no BSM light degrees of freedom matter for the considered observables.}
specific observables can be predicted adapting the most general
WET operator basis, while keeping the initial matching conditions
of the Wilson coefficients at $\muEW$ arbitrary. These
so-called master formulae for various observables can then be
used in any UV completion by simply substituting the
corresponding initial matching conditions. All radiative corrections
below $\muEW$ are properly included, and cancellation of scheme
dependences is guaranteed, provided the initial matching conditions
are determined in the same scheme as the
ADMs.\footnote{
Examples of master formulae can be found in the case of
SMEFT and WET for $\DF$=2 observables in Ref.~\cite{Aebischer:2020dsw},
and for $g-2$ in Ref.~\cite{Aebischer:2021uvt}.}

The perturbative part of this program has been carried out for the
operator basis generated by the SM to very high orders: the initial
Wilson coefficients are known up to the
Next-to-Next-to-Leading Order (NNLO) in QCD~\cite{Bobeth:1999mk}, with
partial NLO electroweak (EW) corrections~\cite{Buras:1999st}. The ADMs
are also calculated to the required precision, including the full NLO QCD
and QED~\cite{Buras:1992tc, Buras:1992zv, Ciuchini:1993vr}, and
partial NNLO QCD contributions~\cite{Gorbahn:2004my}. Further,
also the NLO QCD and QED threshold corrections for $\Nf = 5,4$ are
available~\cite{Buras:1993dy}, as well as the NNLO QCD
ones~\cite{Cerda-Sevilla:2016yzo, Cerda-Sevilla:2018hjk}. As far
as the observable-specific hadronic matrix elements are
concerned, efforts towards their determination have mainly been
focused at the ones that arise in the~SM.

On the other hand, no systematic study exists to similar
precision for the most general $\DF = 1$ WET basis of non-leptonic
operators beyond the SM.  First steps towards the inclusion of NLO
perturbative corrections for non-leptonic $\DF=1,2$ processes in the
presence of new operators have been taken in the literature.  The ADMs
for the complete set of non-leptonic $\DF=1,2$ processes were
calculated at the NLO in QCD in Ref.~\cite{Buras:2000if},
extending an earlier analysis of Ref.~\cite{Ciuchini:1997bw}. The
specific choice~\cite{Buras:2000if} of the operator basis,
called here \emph{BMU basis}, reduced the number of operators that mix
into the QCD- and QED-penguin operators as much as possible, thereby
simplifying the structure of the RG evolution for $\Nf = 3,4,5$. The
BMU basis contains the conventional SM
basis~\cite{Buras:1992tc, Buras:1992zv, Ciuchini:1993vr} as a subset.

In recent years, the idea of a considerable mass gap between
the scale of new physics $\muNP$ and the electroweak scale $\muEW \ll
\muNP$ gave rise to broader interest in the Standard Model
Effective Field Theory (SMEFT)~\cite{Buchmuller:1985jz,
Grzadkowski:2010es}. It assumes the SM field content and the SM
gauge symmetry group $\SUthreeC \otimes \SUtwoL \otimes \UoneY$. In
this context, another WET operator basis has been
introduced~\cite{Jenkins:2017jig} for the tree-level matching of SMEFT
onto WET, called throughout the \emph{JMS basis}.\footnote{In what
follows we use the \texttt{WCxf} convention for the operators defined
in Ref.~\cite{Aebischer:2017ugx}.} This basis has recently been used
in determination of the complete one-loop matching\footnote{
Previous partial results can be found, for
example, in Refs.~\cite{Aebischer:2015fzz, Bobeth:2017xry,
Hurth:2019ula, Endo:2018gdn, Grzadkowski:2008mf}.}
of SMEFT onto WET~\cite{Dekens:2019ept}. Actual
renormalizable UV completions are understood to be matched
onto a stack of effective theories, the last two being
subsequently the SMEFT and the WET.

In this work, we provide a connection of the JMS basis to the
BMU basis beyond the leading order, which requires a correct treatment
of scheme dependences. This combines the advantages of the JMS basis
in the matching to SMEFT with the simplified RG evolution and
known matrix elements in the BMU basis. In
\refsec{sec:nonleptonic-ops} and \refapp{app:non-leptonic-op's},
we recall definitions of the non-leptonic operators in the JMS and BMU bases
for $\Nf = 5$. A transformation of
the Wilson coefficients beyond Leading Order (LO) from the JMS
to the BMU basis is provided in \refsec{sec:basis-change}.
Solutions of the RG equations in the BMU basis
together with a discussion of the $\Nf = 4,3$ cases
are given in \refsec{sec:RG}. In
\refsec{sec:scheme}, we elucidate how the results presented
here should be incorporated into a general SMEFT NLO analysis of
non-leptonic decays. In particular, we indicate how various
non-physical scheme dependences cancel between each other. In
\refsec{sec:numeric}, we provide a numerical analysis for
$\Nf = 5$ that is relevant for non-leptonic $B$ decays, as well as for $\Nf = 3$
that matters for the BSM master formula for $\epe$
in $K\to\pi\pi$ at the NLO in QCD. A summary of our results is
given in \refsec{sec:conclusion}. The relevant ADMs and threshold correction matrices are
collected in the Appendices \ref{app:ADM} and \ref{app:threshold}, respectively.

%
%
%
\section{Non-leptonic operators}
\label{sec:nonleptonic-ops}

We consider in this work non-leptonic $\DF = 1$ transitions of
down-type quarks $d_i \to d_j$ with $i\neq j$ and $i, j \in \{d,s,b\}$.
We focus on $\Delta C = 0$ processes that proceed through
transitions with only the down-type quarks or both the down- and up-type quarks:
\begin{align}
  \label{eq:dddd-op}
  d_i & \to d_j\, d_i \bar d_i ,
& d_i & \to d_j\, d_j \bar d_j ,
& d_i & \to d_j\, d_k \bar d_k \quad (i\neq k\neq j),
\\
  \label{eq:dduu-op}
  d_i & \to d_j\, u_k \bar u_k
& (u_k & = u, c) .
\end{align}
If we did not require $\Delta C = 0$, another type of transition would need to be considered in addition, namely
\begin{align}
  d_i & \to d_j\, u_k \bar u_l ,
& (k & \neq l, \quad k, l \in \{u,c\}) .
\end{align}
A generalization to $\DF = 1$ up-type transitions $c \to u$ is
straightforward.

Below the electroweak scale $\muEW \sim 100\GeV$, these processes are
described within the WET. It allows us to systematically
parameterize effects beyond the SM by including all possible
operators consistent with the local gauge symmetry.
Flavour-changing interactions in the WET Lagrangian that matter for
our purpose are contained in
\begin{align}
  \label{eq:WET-Lag}
  \cL_\text{WET}^\text{4quark} &
  \;=\; \sum_i C_i(\mu) \, \OpL{i}
  \;=\; \vec{C}^T(\mu) \cdot \vec{\OpL{}} ,
\end{align}
where the sum runs over all the relevant four-quark
operators $\OpL{i}$. We are going to restrict our attention to
four-quark operators, leaving aside the chromo- and
electro-magnetic dipole ones. The corresponding Wilson
coefficients $C_i$ depend on the renormalization
scale~$\mu$. They are obtained at the electroweak scale $\mu =
\muEW$ in the matching from either the SMEFT or some of its
possible extensions. As far as the RG evolution to
lower scales is concerned, we prefer to perform it in the BMU basis, see
\refapp{app:def-BMU}. In this basis, the operator mixing is considerably
simplified. Moreover, the ADMs that govern the RG equations are
known already at the NLO in QCD for the complete set of BSM
operators in the BMU basis. Last but not least, calculations of the NLO corrections to
various observables in the literature have been done in this basis.  A
transformation from the JMS to the BMU basis at the NLO in QCD
is given in \refsec{sec:basis-change}.

The non-leptonic operators under consideration are listed in
\refapp{app:non-leptonic-op's}. They can be grouped into six sectors
according to their Dirac structures. The sector names and the
corresponding numbers of operators in each sector are: VLL(8),
VLR(16), SRR(16), VRR(8), VRL(16), SLL(16). Operators from the latter
three sectors (called \emph{chirality-flipped}) are pairwise related
to the former (``non-flipped'') ones via interchange of the chirality
projectors $P_L \leftrightarrow P_R$, where $P_{L,R} \equiv (\One \mp
\gamma_5)/2$. The overall number of operators is $2 \times 40 = 80$
for $\Nf=5$ and $\Delta C = 0$. Since QCD and QED conserve parity, the
full $80\times 80$ ADM exhibits a $2\times 2$ super-block structure,
where each of the super-blocks has dimensions~$40\times 40$. The same
refers to the matrices that describe the JMS$\to$BMU basis
transformation.

We choose the following ordering of the ``non-flipped'' operators in the subsequent sectors of the JMS basis (\refapp{app:def-JMS})
\footnote{In $D=4$, the VLR operators with scalar currents can be re-expressed in terms of vector currents using
the Fierz relations.}
\begin{align}
  \text{VLL} & :
  \big\{ \opL[V1,LL]{ud}{11ji},\; \opL[V8,LL]{ud}{11ji},\;
         \opL[V1,LL]{ud}{22ji},\; \opL[V8,LL]{ud}{22ji},
\notag
\\ & \phantom{: \big\{ \;}
         \opL[V,LL]{dd}{jikk}, \; \opL[V,LL]{dd}{jkki},\;
         \opL[V,LL]{dd}{jiii}, \; \opL[V,LL]{dd}{jijj}
  \big\} , \label{eq:JMS.VLL}
\\
  \text{VLR} & :
  \big\{ \opL[V1,LR]{du}{ji11},\; \opL[V8,LR]{du}{ji11},\;
         \opL[V1,LR]{du}{ji22},\; \opL[V8,LR]{du}{ji22},
\notag
\\ & \phantom{: \big\{ \;}
         \opL[V1,LR]{dd}{jikk}, \; \opL[V8,LR]{dd}{jikk},\;
         \opL[V1,LR]{dd}{jiii}, \; \opL[V8,LR]{dd}{jiii},\;
         \opL[V1,LR]{dd}{jijj}, \; \opL[V8,LR]{dd}{jijj},
\notag
\\ & \phantom{: \big\{ \;}
         \opL[V1,LR]{uddu}{1ji1}^\dagger,\; \opL[V8,LR]{uddu}{1ji1}^\dagger, \;
         \opL[V1,LR]{uddu}{2ji2}^\dagger,\; \opL[V8,LR]{uddu}{2ji2}^\dagger,
\notag
\\ & \phantom{: \big\{ \;}
         \opL[V1,LR]{dd}{jkki}, \; \opL[V8,LR]{dd}{jkki}
  \big\} , \label{eq:JMS.VLR}
\\
  \text{SRR} & :
  \big\{ \opL[S1,RR]{dd}{jiii},\; \opL[S8,RR]{dd}{jiii},\;
         \opL[S1,RR]{dd}{jijj},\; \opL[S8,RR]{dd}{jijj},\;
\notag
\\ & \phantom{: \big\{ \;}
         \opL[S1,RR]{ud}{11ji},\;   \opL[S8,RR]{ud}{11ji},\;
         \opL[S1,RR]{uddu}{1ij1},\; \opL[S8,RR]{uddu}{1ij1},
\notag
\\ & \phantom{: \big\{ \;}
         \opL[S1,RR]{ud}{22ji},\;   \opL[S8,RR]{ud}{22ji},\;
         \opL[S1,RR]{uddu}{2ij2},\; \opL[S8,RR]{uddu}{2ij2},
\notag
\\ & \phantom{: \big\{ \;}
         \opL[S1,RR]{dd}{jikk},\; \opL[S8,RR]{dd}{jikk},\;
         \opL[S1,RR]{dd}{jkki},\; \opL[S8,RR]{dd}{jkki}
  \big\} .  \label{eq:JMS.SRR}
\end{align}
In the chirality-flipped sectors, the ordering takes then the following explicit form:
\begin{align}
  \text{VRR} & :
  \big\{ \opL[V1,RR]{ud}{11ji},\; \opL[V8,RR]{ud}{11ji},\;
         \opL[V1,RR]{ud}{22ji},\; \opL[V8,RR]{ud}{22ji},
\notag
\\ & \phantom{: \big\{ \;}
         \opL[V,RR]{dd}{jikk}, \; \opL[V,RR]{dd}{jkki},\;
         \opL[V,RR]{dd}{jiii}, \; \opL[V,RR]{dd}{jijj}
  \big\} , \label{eq:JMS.VRR}
\\
  \text{VRL} & :
  \big\{  \opL[V1,LR]{ud}{11ji},\; \opL[V8,LR]{ud}{11ji},\;
         \opL[V1,LR]{ud}{22ji},\; \opL[V8,LR]{ud}{22ji},
\notag
\\ & \phantom{: \big\{ \;}
         \opL[V1,LR]{dd}{kkji}, \; \opL[V8,LR]{dd}{kkji},\;
         \opL[V1,LR]{dd}{iiji}, \; \opL[V8,LR]{dd}{iiji},\;
         \opL[V1,LR]{dd}{jjji}, \; \opL[V8,LR]{dd}{jjji},
\notag
\\ & \phantom{: \big\{ \;}
         \opL[V1,LR]{uddu}{1ij1},\; \opL[V8,LR]{uddu}{1ij1}, \;
         \opL[V1,LR]{uddu}{2ij2},\; \opL[V8,LR]{uddu}{2ij2},
\notag
\\ & \phantom{: \big\{ \;}
         \opL[V1,LR]{dd}{kijk}, \; \opL[V8,LR]{dd}{kijk}
  \big\} , \label{eq:JMS.VRL}
\\
  \text{SLL} & :
  \big\{ \opL[S1,RR]{dd}{ijii}^\dagger,\; \opL[S8,RR]{dd}{ijii}^\dagger,\;
         \opL[S1,RR]{dd}{ijjj}^\dagger,\; \opL[S8,RR]{dd}{ijjj}^\dagger,\;
\notag
\\ & \phantom{: \big\{ \;}
         \opL[S1,RR]{ud}{11ij}^\dagger,\;   \opL[S8,RR]{ud}{11ij}^\dagger,\;
         \opL[S1,RR]{uddu}{1ji1}^\dagger,\; \opL[S8,RR]{uddu}{1ji1}^\dagger,
\notag
\\ & \phantom{: \big\{ \;}
         \opL[S1,RR]{ud}{22ij}^\dagger,\;   \opL[S8,RR]{ud}{22ij}^\dagger,\;
         \opL[S1,RR]{uddu}{2ji2}^\dagger,\; \opL[S8,RR]{uddu}{2ji2}^\dagger,
\notag
\\ & \phantom{: \big\{ \;}
         \opL[S1,RR]{dd}{ijkk}^\dagger,\; \opL[S8,RR]{dd}{ijkk}^\dagger,\;
         \opL[S1,RR]{dd}{kjik}^\dagger,\; \opL[S8,RR]{dd}{kjik}^\dagger
  \big\} . \label{eq:JMS.SLL}
\end{align}

Obviously, the BMU basis (\refapp{app:def-BMU}) consists of the same number of operators.
However, the VLL, VLR and the corresponding chirality-flipped sectors get
subdivided into sets that do not mix under the QCD and QED renormalization.
For the purpose of our transformation of the Wilson
coefficients between the JMS and BMU bases in \refsec{sec:basis-change},
we introduce the following reference order:
\begin{align}
  \text{VLL} & :
  \big\{ \OpL{1},\; \OpL{2},\;\;
         \OpL{3},\; \OpL{4},\; \OpL{9},\; \OpL{10},\;\;
         \OpL{11},\;\; \OpL{14}
  \big\} , \label{eq:BMU.VLL}
\\
  \text{VLR} & :
  \big\{ \OpL{5},\; \OpL{6},\; \OpL{7},\; \OpL{8},\;\;
         \OpL{12},\; \OpL{13}, \;\; \OpL{15}, \ldots , \OpL{24}
  \big\} , \label{eq:BMU.VLR}
\\
  \text{SRR} & :
  \big\{ \OpL{25}, \ldots , \OpL{40}
  \big\}. \label{eq:BMU.SRR}
\end{align}

In the chirality-flipped sectors (VRR, VRL and SLL), the ordering is analogous,
up to shifting the operator subscripts ($Q_i \to Q_{40 + i})$.

%
%
%
\section{\bm Basis change JMS$\,\to\,$BMU}
\label{sec:basis-change}

The WET Wilson coefficient RG evolution is most
conveniently performed in the BMU basis for which the corresponding NLO QCD
ADMs are available from Ref.~\cite{Buras:2000if}. Many observables have
already been calculated employing this very
basis. Since the one-loop SMEFT to WET matching has been performed
in the JMS basis, the obtained Wilson coefficients need to be
transformed. At the tree level, it is sufficient to find linear
relations between operators of the two bases in $D=4$ spacetime dimensions.
In our case, they are given by an $80\times 80$ matrix $\hat R$:
\begin{align}
  \label{eq:op-trafo}
  \vec{\OpL{}}_\text{BMU} & = \hat R \, \vec{\OpL{}}_\text{JMS}
& & \Rightarrow
& \vec{C}_\text{BMU} & = (\hat R^{-1})^T \vec{C}_\text{JMS}\;.
\end{align}
Beyond the tree level, the basis choice is a part of the $\overline{\rm MS}$ renormalization scheme definition,
as the treatment of the Dirac algebra in $D\neq 4$ dimensions does matter. In consequence, the Wilson coefficient
transformation receives perturbative NLO corrections of the form
\begin{equation}
  \label{eq:wc-trafo}
\begin{aligned}
  \vec{C}_\text{BMU}(\mu) &
  = (\hat R^{-1})^T \left[ \One
      - \frac{\alS(\mu)}{4\pi} \DelrS^T
      - \frac{\alE(\mu)}{4\pi} \DelrE^T \right] \vec{C}_\text{JMS}(\mu)
\\ &
  \equiv \hat M_\text{JMS}(\mu) \, \vec{C}_\text{JMS}(\mu) .
\end{aligned}
\end{equation}
The Wilson coefficients themselves are determined starting with tree level,
and subsequently adding higher orders in the running couplings $\alS(\mu)$ and $\alE(\mu)$, namely
\begin{align}
  \label{eq:wc-expansion}
  C_i(\mu) &
  = \sum_{j,k=0}^\infty \frac{\alS^j(\mu) \alE^k(\mu)}{(4\pi)^{j+k}} C_i^{(jk)}(\mu)
  = C_i^{(00)}
  + \frac{\alS(\mu)}{4\pi} C_i^{(10)} (\mu)
  + \frac{\alE(\mu)}{4\pi} C_i^{(01)} (\mu) + \ldots \,.
\end{align}
A complete evaluation of the tree-level contributions $C_i^{(00)}$ in the matching of SMEFT onto WET
is known in the JMS basis from Ref.~\cite{Jenkins:2017jig}. It has been extended
to the one-loop level in the SM gauge and Yukawa couplings in
Ref.~\cite{Dekens:2019ept}. The latter calculation thus provides the NLO QCD correction $C_i^{(10)}$,
and the NLO QED correction $C_i^{(01)}$. Effects
from the electroweak gauge boson ($W^\pm, Z^0$) and Higgs exchanges are contained
in $C_i^{(01)}$, too. In consequence, these corrections depend, in general,
on the heavy boson and/or the top-quark masses. In the absence of tree-level contributions in certain cases,
they constitute the first non-vanishing contributions, and higher orders in the perturbative
expansion may become relevant for phenomenological purposes.
We caution the reader that the one-loop
matching results in Ref.~\cite{Dekens:2019ept} are not obtained for the mass-eigenstate basis
-- see details in section 6 of that paper. Hence, they require
an additional rotation of the quark fields to the mass-eigenstate basis.

The matrices $\Delta \hat r_i$ in Eq.~\eqref{eq:wc-trafo} can
be systematically determined in perturbation theory. They
depend on definitions of the physical and evanescent operators in
both bases.  In practice, their determination may require
introducing supplemental evanescent operators to describe the
basis transformation~\cite{Chetyrkin:1997gb}.

One of the complications beyond LO arises due to the Fierz relations required for the
transformation between the JMS and BMU bases. Since they cannot be simply
extended to $D = 4 - 2 \eps$ in the dimensional regularization beyond LO,
Fierz-evanescent operators have to be introduced at the NLO in QCD. The adapted
definitions for these operators in Refs.~\cite{Buras:2000if} and~\cite{Dekens:2019ept}
are equivalent, after accounting for their different colour structures.

Apart from the Fierz-evanescent operators, other evanescent
operators are encountered in Eq.~(4.7) of Ref.~\cite{Dekens:2019ept}. They come
with general ${\mathcal O}(\eps)$ terms that depend on arbitrary constants, of
which only $a_\text{ev}, b_\text{ev}$ and $c_\text{ev}$ are relevant here.
These constants were kept arbitrary in the calculation of the SMEFT
matching conditions for the WET Wilson coefficients in the JMS basis
\cite{Dekens:2019ept}. On the other hand, the corresponding evanescent
operators in the BMU basis~\cite{Buras:2000if}
are defined with the explicit choice $a_\text{ev} = b_\text{ev} = c_\text{ev} = 1$,
which coincides with the so-called Greek projection method~\cite{Tracas:1982gp, Buras:1989xd}.
In principle one could choose to calculate the NLO transformation matrices $\DelrS$ and $\DelrE$
keeping $a_\text{ev}, b_\text{ev}$ and $c_\text{ev}$ arbitrary. Then the
dependence on these constants would cancel between the Wilson coefficients
in the JMS basis and the transformation matrices. However, we refrain from such a complication,
and set $a_\text{ev} = b_\text{ev} = c_\text{ev} = 1$ from the
outset in our calculation of the transformation matrices.
Therefore, also the matched Wilson coefficients in the
JMS basis must be used with this particular choice.

Explicit examples of bases transformations involving Fierz-evanescent
operators at the NLO in QCD are given in Ref.~\cite{Buras:2000if}.
Very similar transformations are encountered
here. Therefore, we shall not describe the calculation in detail but
rather present below our final results only.

The $80\times 80$ tree-level transformation matrix $\hat R$ in Eq.~\eqref{eq:wc-trafo} reads
\begingroup
\setlength\arraycolsep{0.15cm}
\begin{align}
  \hat R & =
  \begin{pmatrix}
    \hat R_\text{VLL} & 0_{8\times 16}    & 0_{8 \times 16} & & &
    \\[0.1cm]
    0_{16\times 8}    & \hat R_\text{VLR} & 0_{16\times 16}
    & & 0_{40\times 40} & \\[0.1cm]
    0_{16\times 8}    & 0_{16\times 16}   & \hat R_\text{SRR}  & & &
    \\[0.1cm]
    & & & \hat R_\text{VLL} & 0_{8\times 16} & 0_{8 \times 16}
    \\[0.1cm]
    & 0_{40\times 40} & &
    0_{16\times 8} & \hat R_\text{VLR}  & 0_{16\times 16}
    \\[0.1cm]
    & & & 0_{16\times 8} & 0_{16\times 16} & \hat R_\text{SRR}
  \end{pmatrix} , \label{eq:rmatrix}
\end{align}
\endgroup
where the upper-left $40 \times 40$ block corresponds to the VLL, VLR
and SRR sectors. The identical lower-right $40 \times 40$ block
transforms the chirality-flipped operators. {The block-diagonal
structure of the matrix $\hat R$ in Eq.~\eqref{eq:rmatrix} is a
consequence of identical ordering that we have chosen for the VLL,
VLR, SRR sectors, as well as for the chirality flipped ones in both
bases -- see Eqs.~\eqref{eq:JMS.VLL}-\eqref{eq:BMU.SRR}. In the BMU
basis case, this ordering does not fully overlap with the original
numbering of operators, as can be seen in
Eqs.~\eqref{eq:BMU.VLL}-\eqref{eq:BMU.SRR}. Each sector undergoes a
separate mapping in the tree-level basis transformation because each
of the relevant Fierz relations involves operators belonging to the
same sector only.} In the individual sectors, {our results for
the sub-blocks of $\hat R$~\eqref{eq:rmatrix} are the following:}
\begin{equation}
  \hat R_\text{VLL} =
  \begin{pmatrix}
  1 & 0 & 0 & 0 & 0 & 0 & 0 & 0 \\[0.1cm]
  \frac{1}{\Nc} & 2 & 0 & 0 & 0 & 0 & 0 & 0 \\[0.1cm]
  1 & 0 & 1 & 0 & 1 & 0 & 1 & 1 \\[0.1cm]
  \frac{1}{\Nc}  & 2 & \frac{1}{\Nc} & 2 & 0 & 1 & 1 & 1 \\[0.1cm]
  1 & 0 & 1 & 0 & -\frac{1}{2} & 0 & -\frac{1}{2} & -\frac{1}{2} \\[0.1cm]
  \frac{1}{\Nc}  & 2 & \frac{1}{\Nc} & 2 & 0 &-\frac{1}{2} &
         -\frac{1}{2}& -\frac{1}{2} \\[0.1cm]
  0 & 0 & 0 & 0 & 0 & 0 & 1 &  1 \\[0.1cm]
  0 & 0 & 0 & 0 & 0 & 0 & 1 & -1
  \end{pmatrix} ,
  \qquad
  \begin{array}{c}
  \hat R_\text{VLR} =
  \begin{pmatrix}
  \hat A_\text{VLR} & 0_{10\times 2}    & 0_{10\times 2}    & 0_{10\times 2} \\[0.1cm]
  0_{2\times 10}    & \hat B_\text{VLR} & 0_{2 \times 2}    & 0_{2 \times 2} \\[0.1cm]
  0_{2\times 10}    & 0_{2 \times 2}    & \hat B_\text{VLR} & 0_{2 \times 2} \\[0.1cm]
  0_{2\times 10}    & 0_{2 \times 2}    & 0_{2 \times 2}    & \hat B_\text{VLR}
  \end{pmatrix},
  \\
  \phantom{X}
  \\
  \hat R_\text{SRR} =
  \begin{pmatrix}
  \hat A_\text{SRR} & 0_{2\times 2}     & 0_{2\times 4}     & 0_{2\times 4}     & 0_{2\times 4} \\[0.1cm]
  0_{2\times 2}     & \hat A_\text{SRR} & 0_{2\times 4}     & 0_{2\times 4}     & 0_{2\times 4} \\[0.1cm]
  0_{4\times 2}     & 0_{4\times 2}     & \hat B_\text{SRR} & 0_{4\times 4}     & 0_{4\times 4} \\[0.1cm]
  0_{4\times 2}     & 0_{4\times 2}     & 0_{4\times 4}     & \hat B_\text{SRR} & 0_{4\times 4} \\[0.1cm]
  0_{4\times 2}     & 0_{4\times 2}     & 0_{4\times 4}     & 0_{4\times 4}     & \hat B_\text{SRR}
  \end{pmatrix} ,
  \end{array}
\end{equation}
with the following blocks:
\begin{equation}
  \hat A_\text{VLR} =
  \begin{pmatrix}
   1 & 0 & 1 & 0 & 1 & 0 & 1 & 0 & 1 & 0 \\[0.1cm]
   \frac{1}{\Nc} & 2 & \frac{1}{\Nc} & 2 & \frac{1}{\Nc} & 2 &
   \frac{1}{\Nc} & 2 & \frac{1}{\Nc} & 2 \\[0.1cm]
   1 & 0 & 1 & 0 & -\frac{1}{2} & 0 & -\frac{1}{2} & 0 & -\frac{1}{2} & 0 \\[0.1cm]
    \frac{1}{\Nc} & 2 & \frac{1}{\Nc} & 2 & -\frac{1}{2 \Nc} & -1 &
   -\frac{1}{2 \Nc} & -1 & -\frac{1}{2\Nc} & -1 \\[0.1cm]
   0 & 0 & 0 & 0 & 0 & 0 & \frac{1}{\Nc} & 2 & \frac{1}{\Nc} & 2 \\[0.1cm]
   0 & 0 & 0 & 0 & 0 & 0 & 1 & 0 & 1 & 0 \\[0.1cm]
   0 & 0 & 0 & 0 & 0 & 0 & \frac{1}{\Nc} & 2 & -\frac{1}{\Nc} & -2 \\[0.1cm]
   0 & 0 & 0 & 0 & 0 & 0 & 1 & 0 & -1 & 0 \\[0.1cm]
   \frac{1}{\Nc} & 2 & -\frac{1}{\Nc} & -2 & 0 & 0 & 0 & 0 & 0 & 0 \\[0.1cm]
   1 & 0 & -1 & 0 & 0 & 0 & 0 & 0 & 0 & 0
  \end{pmatrix} ,
  \qquad
  \begin{array}{c}
  \hat B_\text{VLR} = -\frac{1}{2}
  \begin{pmatrix}
   1 & 0 \\[0.1cm]
   \frac{1}{\Nc} & 2
  \end{pmatrix} ,
  \\
  \phantom{X}
  \\
  \begingroup
  \setlength\arraycolsep{0.15cm}
  \hat A_\text{SRR} =
  \begin{pmatrix}
    1 & 0 \\[0.1cm]
    -\frac{4\Nc+8}{\Nc} & -16
  \end{pmatrix} ,
  \endgroup
  \\
  \phantom{X}
  \\
  \hat B_\text{SRR} =
  \begin{pmatrix}
  \frac{1}{\Nc} & 2 & 0 & 0 \\[0.1cm]
  1 & 0 & 0 & 0 \\[0.1cm]
  -\frac{4}{\Nc} & -8 & -8 & 0 \\[0.1cm]
  -4 & 0 & -\frac{8}{\Nc} & -16
  \end{pmatrix} .
  \end{array}
\end{equation}
The factors $\Nc$ arise from the identity $\delta_{\alpha\delta}
\delta_{\beta\gamma} = \delta_{\alpha\beta} \delta_{\gamma\delta}/\Nc
+ 2\, T^A_{\alpha\beta} T^A_{\gamma\delta}$ for the generators $T^A$
of the $\SUthreeC$ colour algebra.  {As one can see in the above
equations, $\hat R_\text{SRR}$ takes a simple block-diagonal form that
is easy to determine from Fierz identities that relate operators of
the same flavour content. On the other hand, the matrices $\hat
R_\text{VLL}$ and $\hat R_\text{VLR}$ are somewhat more complicated,
which is due to different guiding principles in organizing the JMS and
BMU bases. The JMS one is organized according to the up- and down-type
quark content of the quark currents, while the BMU one is organized
according to the particular pattern of the QCD and QED mixing via
penguin diagrams. In consequence, the QCD and QED penguin operators in
the BMU basis contain (weighted) sums over the up- and down-type quark
currents.}

The NLO QCD correction $\DelrS$ in Eq.~\eqref{eq:wc-trafo} equals to
\begingroup
\setlength\arraycolsep{0.2cm}
\begin{align}
  \DelrS & = \begin{pmatrix}
    \Delta \hat X   & 0_{40 \times 40} \\[0.1cm]
    0_{40\times 40} & \Delta \hat X \\[0.1cm]
  \end{pmatrix} ,
\end{align}
\endgroup
where the $40\times 40$ matrix $\Delta \hat X$ has the following block-diagonal structure:
\begingroup
\setlength\arraycolsep{0.2cm}
\begin{align}
  \label{eq:def-mat-delX}
  \Delta \hat X & = \begin{pmatrix}
    \Delta \hat A  & 0_{8 \times 6} & 0_{8 \times 2} & 0_{8 \times 2} & 0_{8 \times 4} & 0_{8 \times 4} & 0_{8 \times 4} \\[0.1cm]
    0_{16\times 18} & 0_{16\times 6} & 0_{16\times 2} & 0_{16\times 2} & 0_{16\times 4} & 0_{16\times 4} & 0_{16\times 4} \\[0.1cm]
    0_{2\times 18}  & 0_{2 \times 6} & \Delta \hat B  & 0_{2 \times 2} & 0_{2 \times 4} & 0_{2 \times 4} & 0_{2 \times 4} \\[0.1cm]
    0_{2\times 18}  & 0_{2 \times 6} & 0_{2 \times 2} & \Delta \hat B  & 0_{2 \times 4} & 0_{2 \times 4} & 0_{2 \times 4} \\[0.1cm]
    0_{4\times 18}  & 0_{4 \times 6} & 0_{4 \times 2} & 0_{4 \times 2} & \Delta \hat C  & 0_{4 \times 4} & 0_{4 \times 4} \\[0.1cm]
    0_{4\times 18}  & 0_{4 \times 6} & 0_{4 \times 2} & 0_{4 \times 2} & 0_{4 \times 4} & \Delta \hat C  & 0_{4 \times 4} \\[0.1cm]
    0_{4\times 18}  & 0_{4 \times 6} & 0_{4 \times 2} & 0_{4 \times 2} & 0_{4 \times 4} & 0_{4 \times 4} & \Delta \hat C  \\[0.1cm]
  \end{pmatrix} .
\end{align}
\endgroup
Its upper-left $24\times24$ block corresponds to the VLL and VLR operators.
Inside this block, the upper-left $8\times 18$ sub-block $\Delta \hat A$
turns out to be a product of $8 \times 1$ and $1 \times 18$ matrices, namely
\begin{eqnarray}
  \Delta \hat A &=& \frac{1}{6} \;
    \begin{pmatrix}
    0 \quad    1 \quad
    0 \quad   -1 \quad
    0 \quad   -2 \quad
    0 \quad 0
  \end{pmatrix}^T \nonumber\\
  &\times& \Big(
    0 \quad 2 \quad 0 \quad 2 \quad
   -\tfrac{1}{\Nc} \quad 1 \quad \tfrac{\Nc-1}{\Nc} \quad \tfrac{\Nc-1}{\Nc} \quad
    0 \quad 2 \quad 0 \quad 2 \quad 0 \quad 2 \quad 0 \quad 2 \quad 0 \quad 2
    \Big) . \label{eq:DeltaA}
\end{eqnarray}
The remaining lower-right $16\times16$ block of $\Delta \hat X$ corresponds to the SRR sector.
All its non-vanishing entries appear in the diagonal sub-blocks $\Delta \hat B$ and $\Delta \hat C$ whose explicit forms are found to be
\begingroup
\setlength\arraycolsep{0.3cm}
\begin{align}
  \Delta \hat B & = \begin{pmatrix}
    0 & 0 \\[0.1cm]
    -\frac{1}{\Nc^2} - \frac{1}{\Nc} + 1 + \Nc &
    -\frac{2}{\Nc} + 4 + \frac{\Nc}{2}
  \end{pmatrix},
\\
  \Delta \hat C & = \begin{pmatrix}
     0 & 0 & 0 & 0 \\[0.1cm]
     0 & 0 & 0 & 0 \\[0.1cm]
    -\frac{6}{\Nc^2} + 6             & -\frac{12}{\Nc} + 2 \Nc &
    -\frac{1}{2 \Nc} + \frac{\Nc}{2} & 3  \\[0.2cm]
     \frac{3}{\Nc^3} - \frac{4}{\Nc} + \Nc & \frac{6}{\Nc^2} + 3 &
    -\frac{3}{4 \Nc^2} + \frac{3}{4} & -\frac{7}{2\Nc} + \frac{\Nc}{2}
  \end{pmatrix}.
\end{align}
\endgroup
{As we already have mentioned, non-vanishing entries in $\Delta
\hat r_s$ arise whenever the tree-level transformation between the JMS
and BMU bases involves Dirac algebra manipulations that are valid in
$D=4$ dimensions only. In such cases, evanescent operators need to be
taken into account when working out the basis transformation in $D\neq
4$ dimensions. The matrix $\Delta\hat A$ corresponds to the VLL and
VLR sectors taken together. The transformations described by
$\Delta\hat B$ and $\Delta\hat C$ correspond to the SRR-sector
operators with two or three different flavours, respectively. To
determine these matrices, we had to evaluate one-loop diagrams with
insertions of various Fierz-evanescent operators. In the SRR sector
case, no penguin diagrams can give non-vanishing contributions, and
only the current-current diagrams matter. Conversely, no
current-current diagrams with Fierz-evanescent operator insertions
turn out to contribute to $\Delta\hat A$ once our conventions have
been specified as described in the beginning of this section
($a_\text{ev} = b_\text{ev} = c_\text{ev} = 1$). The simple structure
of $\Delta\hat A$ in Eq.~\eqref{eq:DeltaA} stems from the fact that
only penguin diagrams with Fierz-evanescent operator insertions
contribute to this matrix.}

%
%
%
\section{Renormalization group equations}
\label{sec:RG}

The Wilson coefficients satisfy the RG equation
\begin{align}
  \label{eq:wc-rge}
  \mu \frac{d}{d\mu} \vec{C}(\mu) &
  = \adm^T(\mu) \vec{C}(\mu)\,,
\end{align}
governed by the ADM $\adm(\mu)$ of the WET operators. They are renormalized
in the \MSbar{} scheme, implying the proper treatment\footnote{One-loop matrix
elements of evanescent operators can give finite contributions proportional
to tree-level matrix elements of physical operators. Such finite contributions
have to be absorbed into renormalization constants. Although such a finite
renormalization does not fully comply with the notion of ``minimal subtraction'',
this is still referred to as the \MSbar{} scheme. The ADM's depend on these
finite parts. Such scheme dependences cancel at the level of observables
against the corresponding dependences in matrix elements.} of contributions
from evanescent operators~\cite{Buras:1989xd, Dugan:1990df, Herrlich:1994kh}.
The ADM has the following perturbative expansion:
\begin{align}
  \label{eq:adm-exp}
  \adm &
  = \frac{\alS}{4 \pi} \adm^{(10)}
  + \frac{\alS^2}{(4 \pi)^2} \adm^{(20)}
  + \frac{\alE}{4 \pi} \adm^{(01)}
  + \frac{\alE \alS}{(4 \pi)^2} \adm^{(11)}
  + \ldots,
\end{align}
where the ellipses stand for higher-order terms, i.e. $\cO(\alS^3,\,
\alE \alS^2,\, \alE^2)$.

The ADM has a simple block structure in the BMU basis \cite{Buras:2000if}
because many sub-sectors of non-leptonic operators are closed under QCD and
QED renormalization. Operators in such sub-sectors mix only among themselves,
which we refer to as ``self-mixing''. The only exceptions are the sub-sectors
VLL,$u$ (or, alternatively, VLL,$c$), VLL,$i+j$, VLR,$i+j$ and their chirality
flipped counterparts where mixing into the QCD- and QED-penguin operators
occurs in addition. In consequence, off-diagonal blocks arise in the
corresponding parts of the ADM. Focusing on the chirality non-flipped case,
one can write
\begingroup
\setlength\arraycolsep{0.3cm}
\begin{align}
  \label{eq:adm-mix-in-P}
  \adm & \supset
  \begin{pmatrix}
  \adm_{\text{VLL},u} & \adm_{\text{VLL},u   \to P} & 0 & 0  \\[0.2cm]
  0 & \adm_{P} & 0 & 0                                       \\[0.2cm]
  0 & \adm_{\text{VLL},i+j \to P}& \adm_{\text{VLL},i+j} & 0 \\[0.2cm]
  0 & \adm_{\text{VLR},i+j \to P}& 0 & \adm_{\text{VLR},i+j}
  \end{pmatrix} ,
\end{align}
\endgroup
where the diagonal blocks other than $\adm_{P}$ arise from
self-mixing only, while the non-vanishing off-diagonal blocks are due
to mixing into the QCD- and QED-penguin operators.

The one-loop ADMs $\adm^{(10)}$ for QCD have been collected
in Refs.~\cite{Buras:2000if, Aebischer:2017gaw, Aebischer:2018rrz} for the
various sectors of operators. The two-loop QCD ADM $\adm^{(20)}$ is
known for the SM operators from Refs.~\cite{Buras:1992tc, Ciuchini:1993vr}
and for the BSM operators from Ref.~\cite{Buras:2000if}.\footnote{
Non-penguin contributions to the ADMs were first presented
in Ref.~\cite{Ciuchini:1997bw}.} Higher order QCD corrections are
numerically particularly important due to the large value of the
strong coupling $\alS$, especially when it evolves to
lower scales $\mu \lesssim m_b$. Moreover, at the NLO, non-trivial scheme
dependences cancel for the first time.

On the other hand, QED corrections are negligible in general, with exceptions
when the leading QCD contributions cancel in a specific observable. For what
concerns the BSM sub-sectors (SRR,$Q$, SRR,$i(j)$, SRL,$Q$, \ldots), the RG
effects due to QED are generically small thanks to the block structure of the
full ADM. Such a structure guarantees that they are \emph{not} generated via
operator mixing from other sectors under both the QCD and QED RG evolution.
Therefore, in the self-mixing of each BSM sector,\footnote{Except for the
chirality-flipped counterparts of the QED-penguin operators, as they receive
RG-induced corrections from the chirality-flipped VRR,$u$, VRR$,i+j$ and
VRL,$i+j$ operators.} the dominant effect comes from the LO and NLO QCD ADMs
$\adm^{(10)}$ and $\adm^{(20)}$. In this respect, the role of QED-penguin
operators $P_\text{QED}$, see Eq.~\eqref{eq:QED-peng-op}, is special because
they are the only ones that can be generated due to operator mixing at one-loop
in QED from the sub-sectors VLL,$u$, VLL$,i+j$ and VLR,$i+j$, as shown in
Eq.~\eqref{eq:adm-mix-in-P}.
Still, whether the QED contributions $\adm^{(01)}$ and $\adm^{(11)}$ to the
ADM are really required depends on the observable and on the BSM scenario
under consideration. We list therefore these additional matrices only
for those sectors that mix into the QED-penguin operators
as in Eq.~\eqref{eq:adm-mix-in-P}.

In the SM, the QED-penguin operators receive direct matching contributions at
$\muEW$ that are one-loop and $\alE$-suppressed. Therefore, the leading
contributions to their Wilson coefficients at low scales $\muLow$ come actually
from the RG mixing due to VLL,$u$ operators that is given by $\adm^{(01)}$.
Such contributions are enhanced by $\ln(\muEW\,/\,\muLow)$ with respect to
the direct matching contributions. On the other hand, in BSM scenarios that
allow tree-level contributions to the penguin operators, the direct matching
contributions may become relevant, even though they are usually suppressed by
the corresponding NP scale $\muNP$. In such cases, the numerically leading
RG contribution may likely be due to the self-mixing of QED-penguin operators
under the one-loop QCD ADM $\adm^{(10)}$. Whether the SM or
the BSM contributions are numerically leading in the QED-penguin operators depends
then strongly on the numerical values of the NP couplings w.r.t.\ $\alE$, and
has to be checked case by case. We emphasize that this is only of real concern for
observables where the contributions of QED-penguin operators can compete
numerically with the ones of charged-current or QCD-penguin operators,
as, for example, in $\epe$. In view of that the SM matching contributions for
QED-penguin operators are known immediately in the BMU basis,\footnote{This
is true for all the SM matching contributions.} the knowledge of
the matrix $\DelrE$ in Eq.~\eqref{eq:wc-trafo} is only needed for BSM scenarios
where the NP contributions arise as loop-suppressed QED corrections
to guarantee full scheme independence. Since the RG equations are linear, the
SM and BSM contributions can always be evolved separately from each other.

The solution to the RG equation \eqref{eq:wc-rge} with several threshold
crossings from $\Nf=5$ to $\Nf=3$ active quark flavours has the following general
form:
\begin{equation}
  \label{eq:rge-sol-with-thr}
\begin{aligned}
  \vec{C}^{(3)}(\mu) &
  = \hat U^{(3)}(\mu, \mu_4) \; \hat M^{(4)}(\mu_4)
\\ &
  \times \hat U^{(4)}(\mu_4, \mu_5) \; \hat M^{(5)}(\mu_5)
\\ &
  \times \hat U^{(5)}(\mu_5, \muEW) \; \vec{C}^{(5)}(\muEW).
\end{aligned}
\end{equation}
Here the scales are ordered as $\mu_4 < \mu_5 < \muEW$, and the
threshold scales are comparable to the masses of the decoupled quarks:
$\mu_5 \sim m_b$ and $\mu_4 \sim m_c$. The Wilson coefficients
$\vec{C}^{(\Nf)}_i$ are vectors with dimensions that correspond to the
number of operators present in the effective theories. The matrices
$\hat M^{(\Nf)}$ account for threshold corrections that arise when going from the $\Nf$- to
the $(\Nf-1)$-flavour WET, namely
\begin{equation}
  \label{eq:def-thr-corr}
\begin{aligned}
  \vec{C}^{(\Nf-1)} (\mu_{\Nf}) &
  = \hat M^{(\Nf)}(\mu_{\Nf})\; \vec{C}^{(\Nf)}(\mu_{\Nf}).
\end{aligned}
\end{equation}
They do not need to be quadratic matrices if the
number of operators changes at the threshold.
They are found by calculating partonic matrix elements of operators
in the $\Nf$-flavour WET. In evaluating such matrix elements, we treat the $\Nf$-th quark as
massive with mass $m_{\Nf}$, and all other quarks as massless. Scheme dependences
cancel in the product with the Wilson coefficients. Finally, the RG
running of the Wilson coefficients is accounted for by the evolution operators $\hat U^{(\Nf)}$
\begin{align}
  \label{eq:evol-RGE-gen}
  \vec{C}^{(\Nf)} (\mu_{\Nf}) &
  = \hat U^{(\Nf)}(\mu_{\Nf}, \mu_{\Nf+1}) \;
    \vec{C}^{(\Nf)}(\mu_{\Nf+1})
\end{align}
that multiply the initial conditions
$\vec{C}^{(\Nf)}(\mu_{\Nf+1})$ at the scale $\mu_{\Nf+1} >
\mu_{\Nf}$. The evolution operators depend on the ADMs
$\adm^{(\Nf)}$ of the $\Nf$-flavour WET, as well as on the
running couplings $\alS^{(\Nf)}$ and $\alE^{(\Nf)}$.
By solving the RG equations, one finds that the evolution operator for fixed $\Nf$ and
including the LO and NLO QCD contributions is given by
\begin{align}
  \label{eq:RG-U}
  \hat U^{(\Nf)}(\mu_1, \mu_2) &
  = V \left( D  + \frac{\alS(\mu_1)}{4\pi} H \right) V^{-1}
    + \ldots
\end{align}
 where the ellipses indicate higher orders in $\alS$ and $\alE$. The matrix
$V$ diagonalizes the transposed LO ADM $\adm^{(10)}$. To describe the
matrices $D$ and $H$ in Eq.~\eqref{eq:RG-U}, we begin with defining
\begin{align}
  G^{(0)} & = V^{-1}\, \left(\adm^{(10)} \right)^T\, V ,
&
  G^{(1)} & = V^{-1}\, \left(\adm^{(20)} \right)^T\, V ,
\end{align}
where $G^{(0)}$ is diagonal, while $G^{(1)}$ usually contains off-diagonal terms.
Then $D$ is given by the diagonal matrix
\begin{align}
  D_{ij} & = \delta_{ij} \, \eta^{\,G^{(0)}_{ii} / (2\, \beta_0)},
\hspace{2cm} \mbox{with}
& \eta & = \frac{\alS(\mu_2)}{\alS(\mu_1)},
\end{align}
while the expression for $H$ reads
\begin{align} \label{eq:Hmatrix}
  H_{ij} & =
    \left[ G^{(1)} - G^{(0)} \frac{\beta_1}{\beta_0}\right]_{ij}
    \times \left\{ \begin{array}{ccc}
       \displaystyle
       \frac{\eta D_{ii}}{2\beta_0} \, \ln \eta,
       & \text{ for } &
       2\beta_0 + G^{(0)}_{ii} - G^{(0)}_{jj} = 0, \\[0.6cm]
       \displaystyle
       \frac{\eta D_{ii} - D_{jj}}{2\beta_0 + G^{(0)}_{ii} - G^{(0)}_{jj}},
       & & \text{otherwise.}
    \end{array} \right.
\end{align}
Here, $\beta_0$ and $\beta_1$ are the first and second coefficients of the beta function for $\alS$,
respectively. For processes/observables where the QED-penguin
operators cannot be neglected, the above expressions need
to be supplemented with non-leading QED corrections. Analytic
solutions to the RG equations in such cases can be found in
Refs.~\cite{Buras:1993dy,Huber:2005ig}.

The LO and NLO ADMs for QCD are collected in \refapp{app:ADM} for
arbitrary $\Nf$. As most of the sub-sectors of operators
have only self-mixing,\footnote{
The entries in the ADMs correspond to the operator ordering in the BMU basis
as given in \refapp{app:def-BMU}.}
the RG evolution in such cases can be performed for each
sub-sector individually, except for the ones indicated in
Eq.~\eqref{eq:adm-mix-in-P}. In \refapp{app:ADM} we also provide
the LO QED and the NLO QED$\times$QCD ADMs, i.e. $\adm^{(01)}$ and $\adm^{(11)}$, respectively.

In the remainder of this section, we discuss linear relations that arise
among the BMU operators when going from $\Nf = 5$ to $\Nf = 4$,
and further to $\Nf = 3$. In practice, they are only relevant for Kaon
physics where the convention is usually $d_i = d$ and $d_j = s$.\footnote{We
adopt here the most popular choice for incoming and outgoing flavours.
Alternatively, $d_i = s$ and $d_j = d$ is sometimes chosen, together with
a complex conjugation of the Wilson coefficients.} These
relations reduce the number of linearly independent operators in the corresponding WETs.
Some of the remaining operator Wilson coefficients are then
affected by threshold corrections that arise already at the tree
level. Such effects were taken into account in Refs.~\cite{Aebischer:2018quc,Aebischer:2018csl}
where the LO BSM master formula for $\epe$ was derived.

When the considered linear relations involve
Fierz identities, a careful treatment beyond the LO is necessary,
with inclusion of the proper evanescent operators. It turns out that
all such cases involve the QCD- and QED-penguin operators.
In this respect, it is a common practice in the
literature to work for $\Nf=3,4$ with a redundant operator set
in the framework of the SM~\cite{Buras:1993dy}, retaining the analogues
of all the QCD- and QED-penguin operators that were
present for $\Nf = 5$. In the next sections, we will comment
on the consequences of such an approach for the ADMs and
threshold corrections.

At the $b$-quark decoupling scale $\mu_5$, we remove operators
that mediate $d_i \to d_j \,b\bar b$. There are six operators in the
two sub-sectors SRR,$b$ (4) and SRL,$b$ (2) that, in
principle, can give rise to threshold corrections to the
QCD- and QED-penguin operators.  Furthermore, in the
QCD- and QED-penguin operators $\OpL{3,\ldots,10}$, the sum
runs now only over $q = u,d,s,c$. In consequence, some of
our operators become linearly dependent:
\begin{align}
  \label{eq:nf4-1}
  \OpL{10} & = \OpL{9} + \OpL{4} - \OpL{3} + \boxed{F} ,
\\
  \label{eq:nf4-2}
  \OpL{11} & = \frac{2}{3} (\OpL{3} - \OpL{9}) ,
\\
  \label{eq:nf4-3}
  \OpL{12} & = \frac{2}{3} (\OpL{6} - \OpL{8}) ,
&
  \OpL{13} & = \frac{2}{3} (\OpL{5} - \OpL{7}) ,
\end{align}
where $\boxed{F}$ denotes a Fierz-evanescent operator that
requires a careful treatment beyond the LO~\cite{Buras:1993dy}.
In the SM, as mentioned before,
it is a common practice to retain $\OpL{10}$ and work with a redundant set of operators.
The relations in the VLL,$i+j$ and VLR,$i+j$ sub-sectors are valid also
in $D \neq 4$ dimensions, hence without any complication due to evanescent
operators. Furthermore, the elimination of the VLL,$i+j$ and VLR,$i+j$ sub-sectors
removes them as sources of mixing into the QCD- and QED-penguin operators,
see \refsec{sec:RG}. The operators VLL,$i+j$ and VLR,$i+j$ simply give
tree-level threshold corrections to the involved QCD- and QED-penguin operators.
This gives a basis for $\Nf = 4$ with $40 - 10 = 30$ operators
in the ``non-flipped'' sector, and analogously in the chirality-flipped
one. There are 31 operators in the redundant set though when $\OpL{10}$ is retained.

At the $c$-quark decoupling scale $\mu_4$, we remove operators
that mediate $d_i \to d_j \,c\bar c$. There are six
operators in the two sub-sectors SRR,$c$ (4) and SRL,$c$ (2). Furthermore, in
the QCD- and QED-penguin operators, the sum runs now only over
$q = u,d,s$. In effect, two of the QCD- and QED-penguin operators become linearly dependent:
\begin{align}
  \label{eq:nf3-1}
  \OpL{4} & = \OpL{3} + \OpL{2} - \OpL{1} + \boxed{F} ,
\\
  \label{eq:nf3-2}
  \OpL{9} & = \frac{1}{2} (3 \OpL{1} - \OpL{3}) + \boxed{F} .
\end{align}
In the SM, it is common to retain both operators and work with a
redundant set.  {We note that $\boxed{F}$ in
Eqs.~\eqref{eq:nf4-1}, \eqref{eq:nf3-1} and \eqref{eq:nf3-2} stands for
different evanescent operators. They are actually defined by these
equations up to arbitrary $\cO(\epsilon)$ terms, where $\epsilon
= (D-4)/2$. They are not relevant in the present paper, as we choose to work with
a redundant operator set for the SM operators below the $b$-quark
decoupling scale.}

The operators $\OpL{17,18}$ of the VLR,$u-c$ sub-sector reduce to VLR operators
with only up-quarks that can be expressed in terms of the SM penguin operators
\begin{align}
  \label{eq:nf3-3}
  \OpL{17} \;\to\; \OpL[\text{VLR},u]{1} & = \frac{1}{3} (\OpL{6} + 2 \OpL{8}) ,
&
  \OpL{18} \;\to\; \OpL[\text{VLR},u]{2} & = \frac{1}{3} (\OpL{5} + 2 \OpL{7}) .
\end{align}
Both operators give rise to tree-level threshold corrections to the
QCD- and QED-penguin operators.
This gives additional $4+4+2$ operators to be discarded, such that for
$\Nf = 3$ the operator basis contains $30 - 10 = 20$ operators
in the ``non-flipped'' sector, and analogously in the chirality-flipped
one. There are 23 operators though when $\OpL{4,9,10}$ are retained.

%
%
%
\section{Comments on scheme cancellations}
\label{sec:scheme}

It is well known that the two-loop anomalous dimensions depend
on the renormalization scheme (RS) for the effective theory fields
and parameters, including the Wilson coefficients. There is a
correlation between RS dependences of the Wilson coefficients and
matrix elements of the corresponding operators. In dimensional
regularization, the $\overline{\rm MS}$ scheme definition includes
specifying the necessary evanescent operators, as well as choosing
particular Dirac structures in the physical operator
basis~\cite{Buras:2020xsm}. The reader is already familiar with these
issues after studying our paper up to this point. The final physical
amplitudes
\begin{align}
  \mathcal{A} &
  = \langle \vec{\OpL{}}_\text{BMU}(\muLow)\rangle^T \,
    \vec{C}_\text{BMU}(\muLow)
\end{align}
are RS-independent. In particular, they do not depend on definitions of
evanescent operators. Such dependences are cancelled between
$\langle \vec{\OpL{}}_\text{BMU} (\muLow) \rangle$ and
$\vec{C}_\text{BMU}(\muLow)$. The latter Wilson coefficients can be
expressed in terms of the SMEFT ones at the new physics
scale $\muNP$ as follows:
\begin{align}
  \label{eq:fullEvol}
  \vec{C}_\text{BMU}(\muLow) &
  = \hat U_\text{BMU}(\muLow, \muEW) \;
    \hat M_\text{JMS}(\muEW)\, \hat K(\muEW) \;
    \hat U_\text{SMEFT}(\muEW,\muNP) \;
    \vec{\mathcal{C}}_\text{SMEFT}(\muNP) .
\end{align}
Here, the matrix $\hat M_\text{JMS}$ summarizes the JMS$\to$BMU basis transformation
in the WET as given in Eq.~\eqref{eq:wc-trafo}.
The matrix $\hat K$ summarizes matching relations
between the WET in the JMS basis and the SMEFT in the Warsaw basis~\cite{Grzadkowski:2010es}.
It has been calculated including one-loop contributions in Ref.~\cite{Dekens:2019ept}. Explicitly
\begin{align}
  \label{eq:SMEFTWET}
  \vec{\mathcal{C}}_\text{BMU}(\muEW) &
  = \hat M_\text{JMS}(\muEW) \; \vec{\mathcal{C}}_\text{JMS}(\muEW),
&
  \vec{\mathcal{C}}_\text{JMS}(\muEW) &
  = \hat K(\muEW) \; \vec{\mathcal{C}}_\text{SMEFT}(\muEW)\,.
\end{align}
Various contributions to Eq.~\eqref{eq:fullEvol} must be evaluated using
the same RS to guarantee proper cancellation of the RS dependences.
To describe it in more detail, let us factorize out the NLO contributions
to the QCD evolution matrices\footnote{For simplicity, only the RG running
due to QCD interactions is discussed here.}
\begin{align}
  \label{eq:UBMU}
  \hat U_\text{BMU}(\muLow,\, \muEW) &
  = \left[1 + \hat J_\text{BMU}\frac{\alS(\muLow)}{4\pi} \right]
    \hat U_\text{BMU}^{(0)}(\muLow,\, \muEW)
    \left[1 - \hat J_\text{BMU} \frac{\alS(\muEW)}{4\pi} \right],
\\
  \label{eq:USMEFT}
  \hat U_\text{SMEFT}(\muEW,\, \muNP) &
  = \left[1 + \hat J_\text{SMEFT} \frac{\alS(\muEW)}{4\pi} \right]
    \hat U_\text{SMEFT}^{(0)}(\muEW,\, \muNP)
    \left[1 - \hat J_\text{SMEFT} \frac{\alS(\muNP)}{4\pi} \right],
\end{align}
where $\hat U_i^{(0)}$ are the RS-independent LO evolution
matrices. On the other hand, $\hat J_i$ stem from the
RS-dependent two-loop ADMs, which makes them sensitive to the
evanescent operator definitions. Explicit general expressions for
$\hat U_i^{(0)}$ and $\hat J_i$ can be found in Ref.~\cite{Buras:2020xsm}.

Perturbative expansions of $\hat M_\text{JMS}(\muEW)$, $\hat K(\muEW)$ and $\vec{\mathcal{C}}_\text{SMEFT}(\muNP)$ take the form
\begin{align}
  \hat M_\text{JMS}(\muEW) &
  = \hat M^{(0)} +\frac{\alS(\muEW)}{4\pi} \hat M^{(1)},
&
  \hat K(\muEW) &
  = \hat K^{(0)} +\frac{\alS(\muEW)}{4\pi} \hat K^{(1)},
\end{align}
and
\begin{align}
  \vec{\mathcal{C}}_\text{SMEFT}(\muNP) &
  = \vec{\mathcal{C}}^{(0)}_\text{SMEFT} +
    \frac{\alS(\muNP)}{4\pi} \vec{\mathcal{C}}_\text{SMEFT}^{(1)}\,.
\end{align}
Having all the above expressions at hand, we can easily trace out cancellations
of the RS dependences:
\begin{itemize}
\item
  The RS dependence of $\vec{\mathcal{C}}_\text{SMEFT}^{(1)}$ is
  cancelled by the one in $\hat J_\text{SMEFT}$ entering the last factor
  on the r.h.s.\ of Eq.~\eqref{eq:USMEFT}.
\item
  The one in $\hat J_{\rm BMU}$ entering the first factor on the
  r.h.s.\ of Eq.~\eqref{eq:UBMU} is cancelled by the one in
  $\langle\vec{\mathcal{O}}_\text{BMU}(\muLow)\rangle$.
\item
  Finally, the remaining RS dependences, including those related to the
  evanescent operator definitions cancel in the product
  \begin{align*}
    \left[1 - \hat J_{\rm BMU} \frac{\alS(\muEW)}{4\pi} \right]
    \left[\hat M^{(0)} +\frac{\alS(\muEW)}{4\pi}\hat M^{(1)} \right]
    \left[\hat K^{(0)} +\frac{\alS(\muEW)}{4\pi}\hat K^{(1)} \right]
    \left[1 + \hat J_{\rm SMEFT}\frac{\alS(\muEW)}{4\pi} \right] .
  \end{align*}
\end{itemize}
In order to obtain RS-independent results, one has to ensure
that evaluation of the matching matrices $\hat M_\text{JMS}$
and $\hat K$ is consistent with the scheme dependences in $\hat
J_\text{BMU}$ and $\hat J_\text{SMEFT}$. This indeed has been done for
$\hat M_\text{JMS}$ in the present paper.  As far as $\hat K$
in Ref.~\cite{Dekens:2019ept} is concerned, using
precisely the same RS is recommended in the future calculation of
$\hat J_\text{SMEFT}$. Further details on RS dependences can
be found in Refs.~\cite{Buras:2000if, Chetyrkin:1997gb,
Gorbahn:2004my}.

%
%
%
\section{Numerical analysis}
\label{sec:numeric}

In this section, we study the impact of NLO QCD corrections on the four-quark
operators in the following two cases:
\begin{itemize}
\item non-leptonic $B$ decays governed by $b\to d$ and $b\to s$ transitions,
\item the decays $K\to \pi \pi$ governed by $s\to d$ transitions.
\end{itemize}

The central quantities are the RG evolution operators connecting the WET Wilson
coefficients in the BMU basis at low-energy scales $\muLow$ to the ones of
the JMS basis at the electroweak scale $\muEW$
\begin{align}
  \label{eq:def-C_BMU}
  \vec{C}_\text{BMU}(\muLow)
  \; = \; \hat U_\text{BMU}(\muLow, \muEW) \;
    \hat M_\text{JMS}(\muEW)\, \vec{C}_\text{JMS}(\muEW) .
\end{align}
The basis transformation from the JMS to the BMU basis $\hat M_\text{JMS}$ has
been outlined in \refsec{sec:basis-change}.
The exact form of the evolution operator $\hat U_\text{BMU}(\muLow, \muEW)$
depends on the number of crossed quark thresholds. The most general form
required for our purposes is given in Eq.~\eqref{eq:rge-sol-with-thr}.
We study here the elements
\begin{align}
  \label{eq:def-u}
  u_{ab}(\muLow,\muEW) &
  \; \equiv \; \big[ \hat U_\text{BMU}(\muLow, \muEW) \;
                        \hat M_\text{JMS}(\muEW) \big]_{ab} \,
\end{align}
that depend explicitly on the two scales $\muEW$ and
$\muLow$. In the $s\to d$ case, there is also an implicit
dependence on the quark threshold scales $\mu_5$ and $\mu_4$ that
cancels out up to residual higher-order effects. Apart from that,
$u_{ab}(\muLow,\muEW)$ depend only on the gauge coupling constant initial values
at $\mu = m_Z = 91.1876(21) \GeV$~\cite{Zyla:2020zbs}
\begin{align}
   \alS^{(5)}(m_Z) & = 0.1179(10) ,
&
   \alE^{(5)}(m_Z) & = 1/127.952(9),
\end{align}
as well as on the solutions to the RG equations for these couplings, mainly $\alS$.
Throughout our analysis, the RG evolution of $\alS$ is performed at the 3-loop level, which means
that partial beyond-NLO effects are included, too. Furthermore, the \MSbar{} quark
mass values~\cite{Zyla:2020zbs}
\begin{align}
  \oL{m}_c(\oL{m}_c) & = 1.27(2) \GeV ,
&
  \oL{m}_b(\oL{m}_b) & = 4.18(3) \GeV ,
\end{align}
are used in the threshold corrections collected in \refapp{app:threshold}.
We note that the renormalization scheme of the quark masses plays no role
at the NLO level in QCD. The scales for the threshold crossings
are set to
\begin{align}
  \mu_4 & = 1.3 \GeV ,
&
  \mu_5 & = 4.2 \GeV .
\end{align}
The quantities $u_{ab}(\muLow,\muEW)$ are found by solving the general RG equation \eqref{eq:wc-rge}
for the Wilson coefficients. We consider two options:
\begin{enumerate}
\item A perturbative {\em analytic solution} at the NLO in QCD, as outlined in Eqs.~\eqref{eq:RG-U}-\eqref{eq:Hmatrix}.
      See also Refs.~\cite{Buras:1993dy} and \cite{Huber:2005ig} for the coupled QCD$\times$QED case.
\item A direct {\em numerical solution} to the system of ordinary differential
      equations \eqref{eq:wc-rge} together with the RG equation for the couplings $\alS$ and $\alE$.
\end{enumerate}
Both options are equivalent up to higher order corrections but numerically they differ by residual effects.
Furthermore, numerical ambiguities of similar origin can arise from the treatment of products of
quantities that have perturbative expansions in $\alS$ and $\alE$. For example,
products like \eqref{eq:rge-sol-with-thr} and \eqref{eq:def-u} can either be
expanded, thereby discarding higher-order contributions beyond the NLO, or
kept unexpanded, thereby including partial NNLO contributions.
Both results are equivalent at the NLO level. Numerical differences between them can
be used to estimate uncertainties that arise due to lacking NNLO contributions.

Our default choice is the numerical solution to the RG equation, unless stated
otherwise. Products of the RG evolution operators $\hat U^{(\Nf)}$ and threshold
corrections $\hat M^{(\Nf)}$ are kept unexpanded.

%
%
\subsection[Non-leptonic $B$ decays]
{\bm Non-leptonic $B$ decays}
\label{sec:hadr-b-decays}

Hadronic $B$ decays are governed by the four-quark operators specified in
Eq.~\eqref{eq:dddd-op} and \eqref{eq:dduu-op}, with the flavour indices $i = b$
and $j = d$ or $j = s$. In the absence of quark flavour thresholds, the
quantities of interest are
\begin{align}
  \label{eq:def-u-nf5}
  u_{ab}(\muLow,\muEW) & \; \equiv \; \big[ \hat U^{(5)}(\muLow, \muEW) \;
                        \hat M_\text{JMS}(\muEW) \big]_{ab} \, .
\end{align}
The index $b$ runs over all the $2 \times 40$ operators of the JMS basis,
whereas $a$ corresponds to all the $2 \times 40$ operators of the BMU basis
in the $\Nf = 5$ theory, as described in \refsec{sec:nonleptonic-ops} and
\refapp{app:non-leptonic-op's}. The matrices $u_{ab}(\muLow,\muEW)$ consist
of two $40 \times 40$ identical diagonal blocks corresponding the non-flipped
and chirality-flipped operators. Many entries vanish due to the block structure
of the matrix $\hat M_\text{JMS}$, as well as of the ADMs.


\begin{figure}
\centering
  \includegraphics[width=0.238\textwidth]{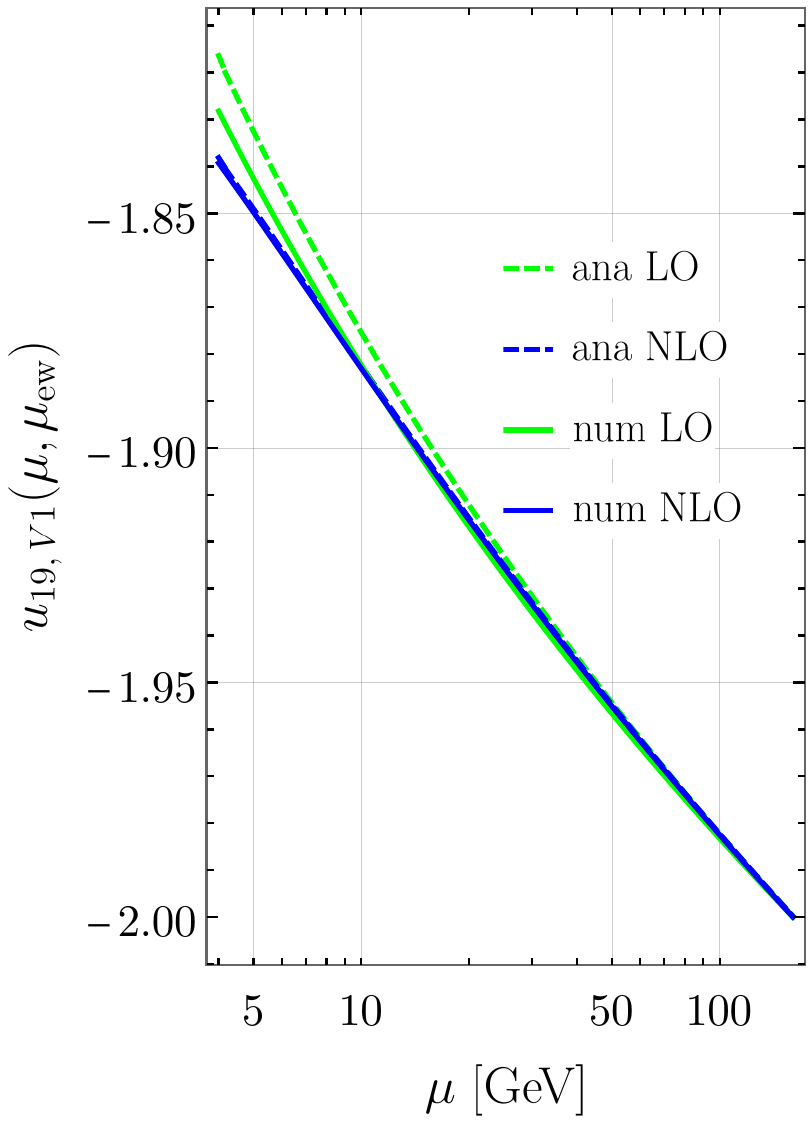}
  \hskip 0.01\textwidth
  \includegraphics[width=0.23\textwidth]{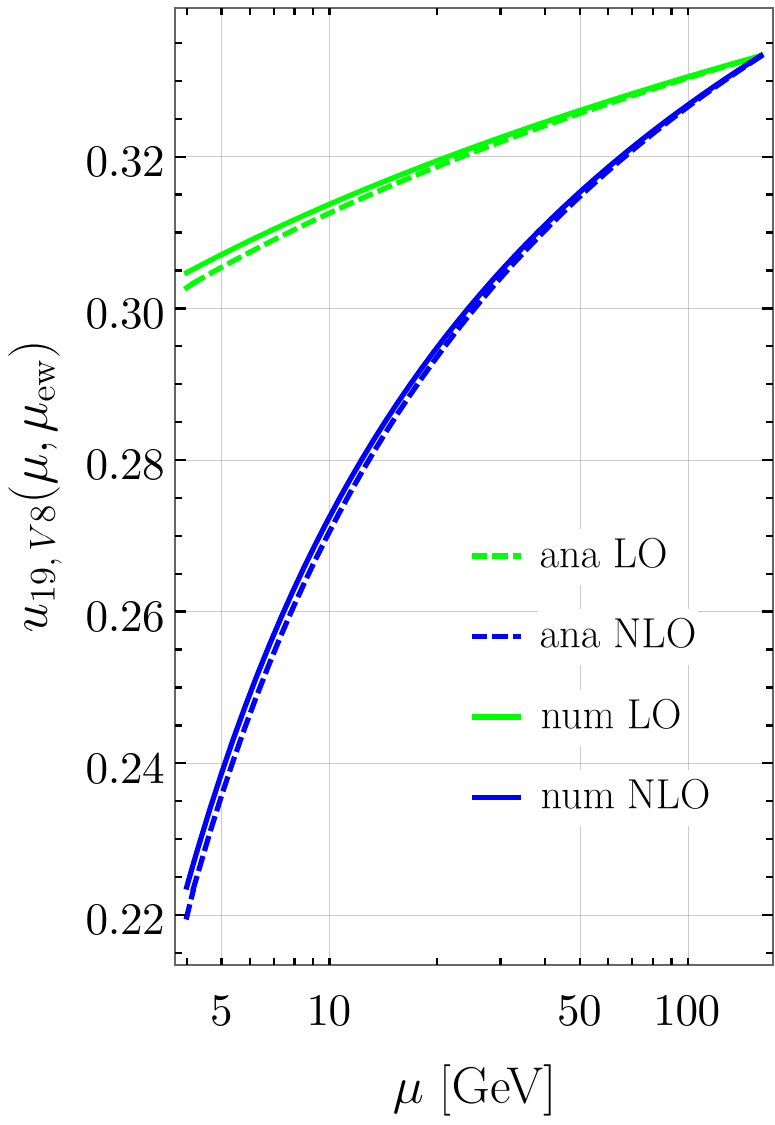}
  \hskip 0.01\textwidth
  \includegraphics[width=0.23\textwidth]{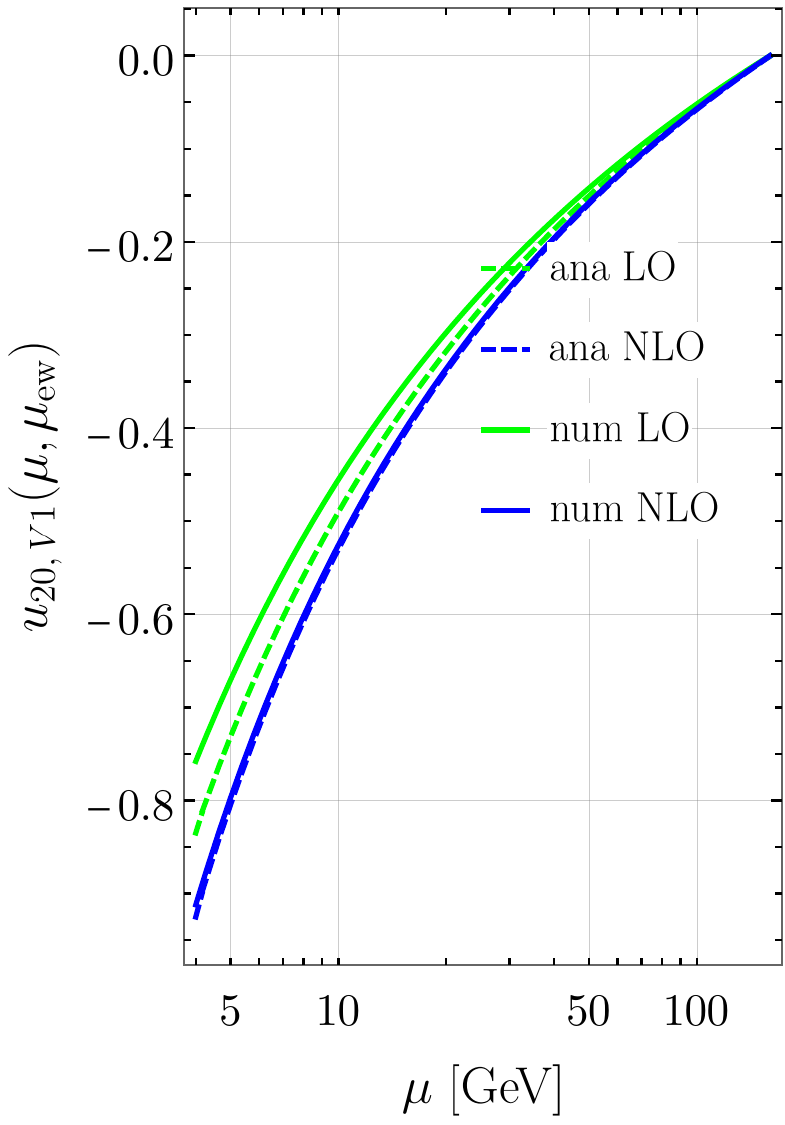}
  \hskip 0.01\textwidth
  \includegraphics[width=0.23\textwidth]{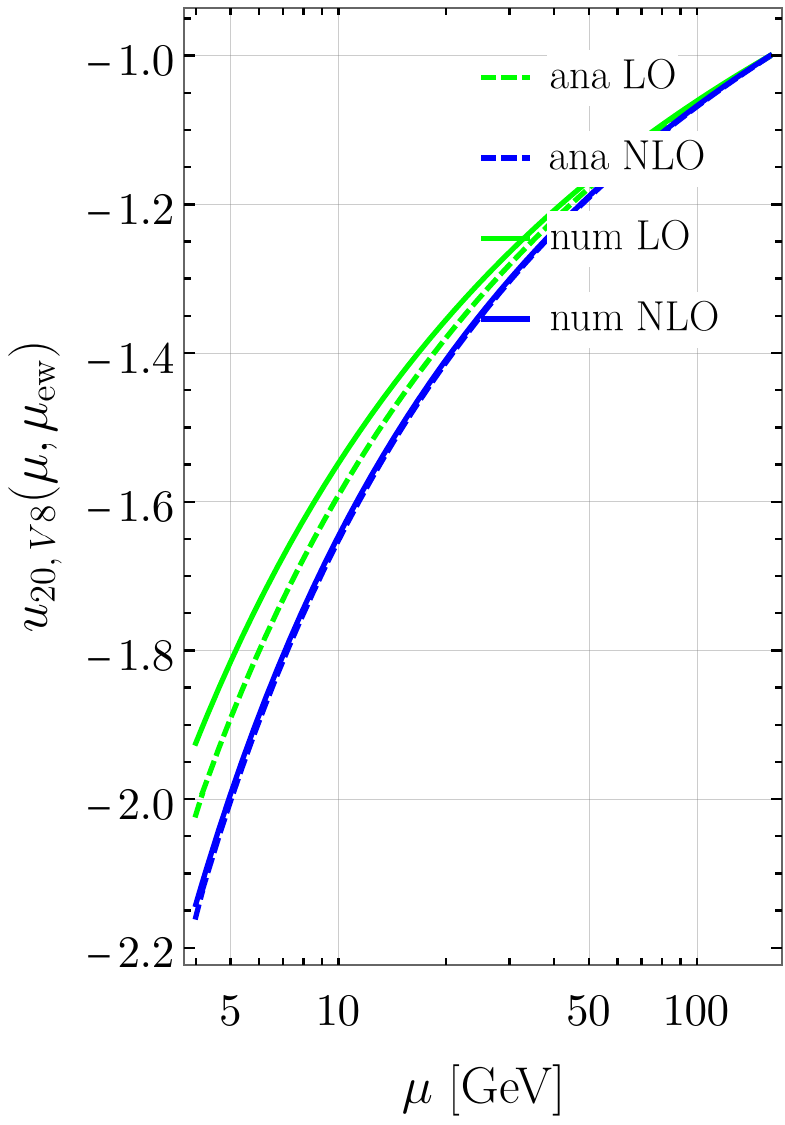}
\\
  \includegraphics[width=0.22\textwidth]{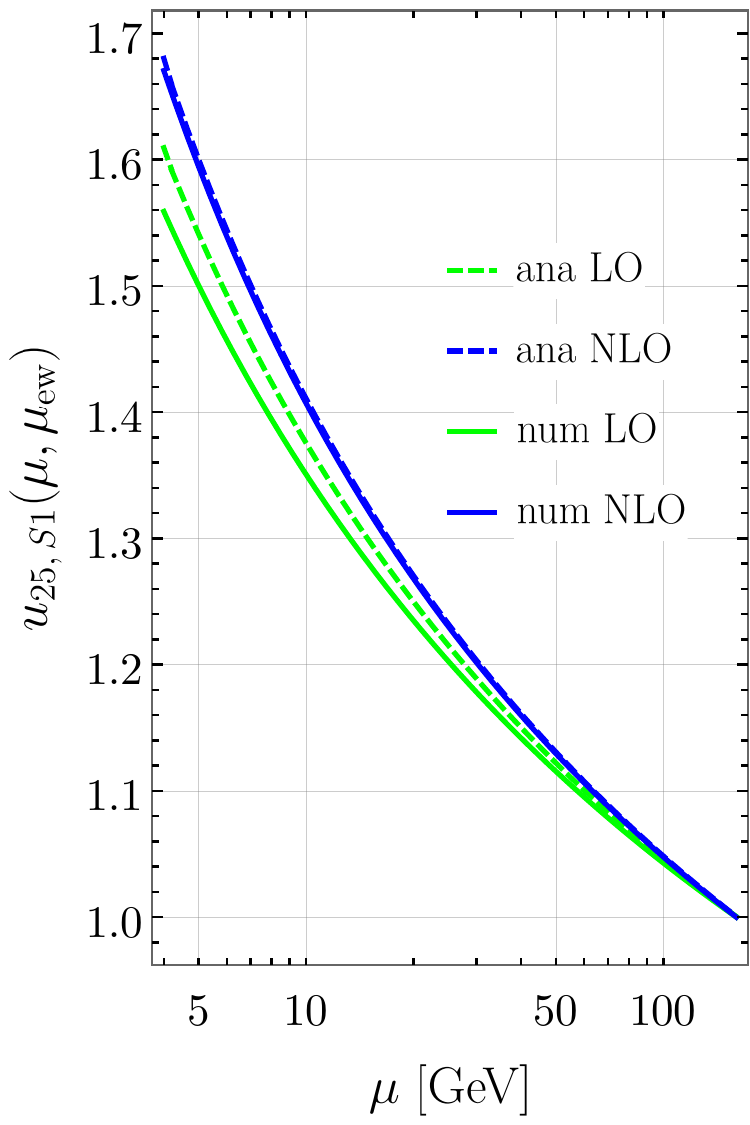}
  \hskip 0.01\textwidth
  \includegraphics[width=0.235\textwidth]{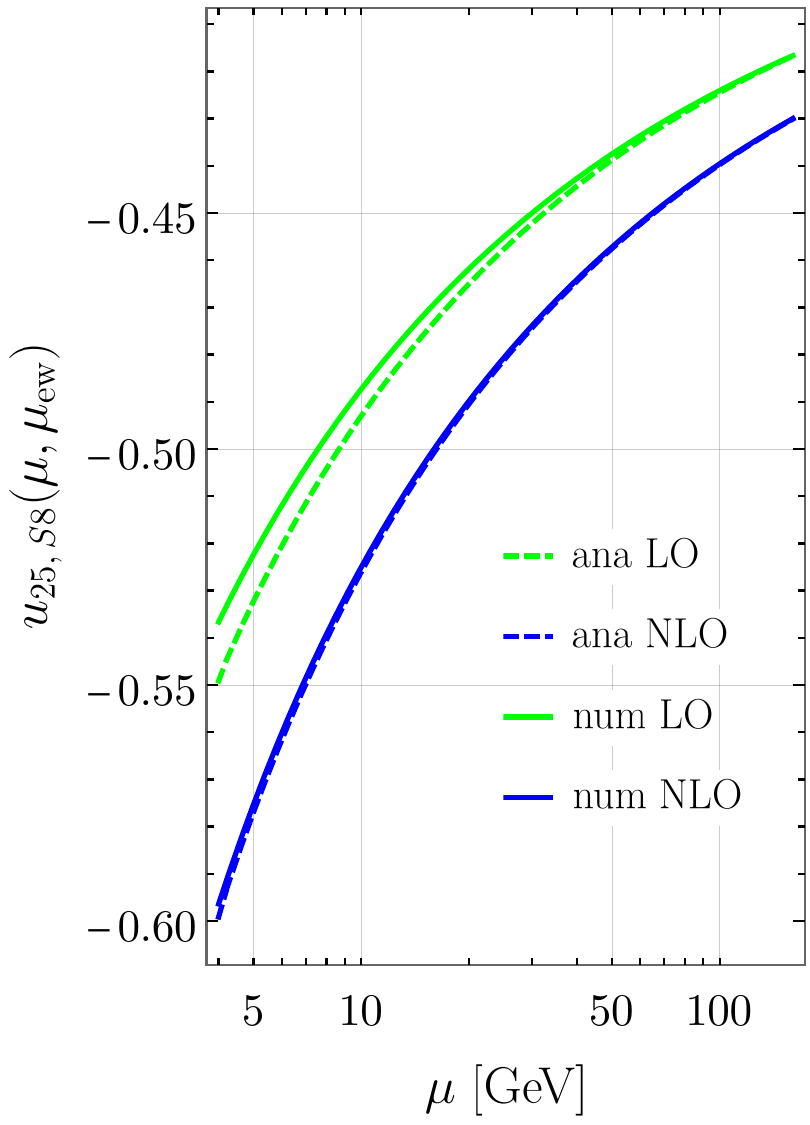}
  \hskip 0.01\textwidth
  \includegraphics[width=0.23\textwidth]{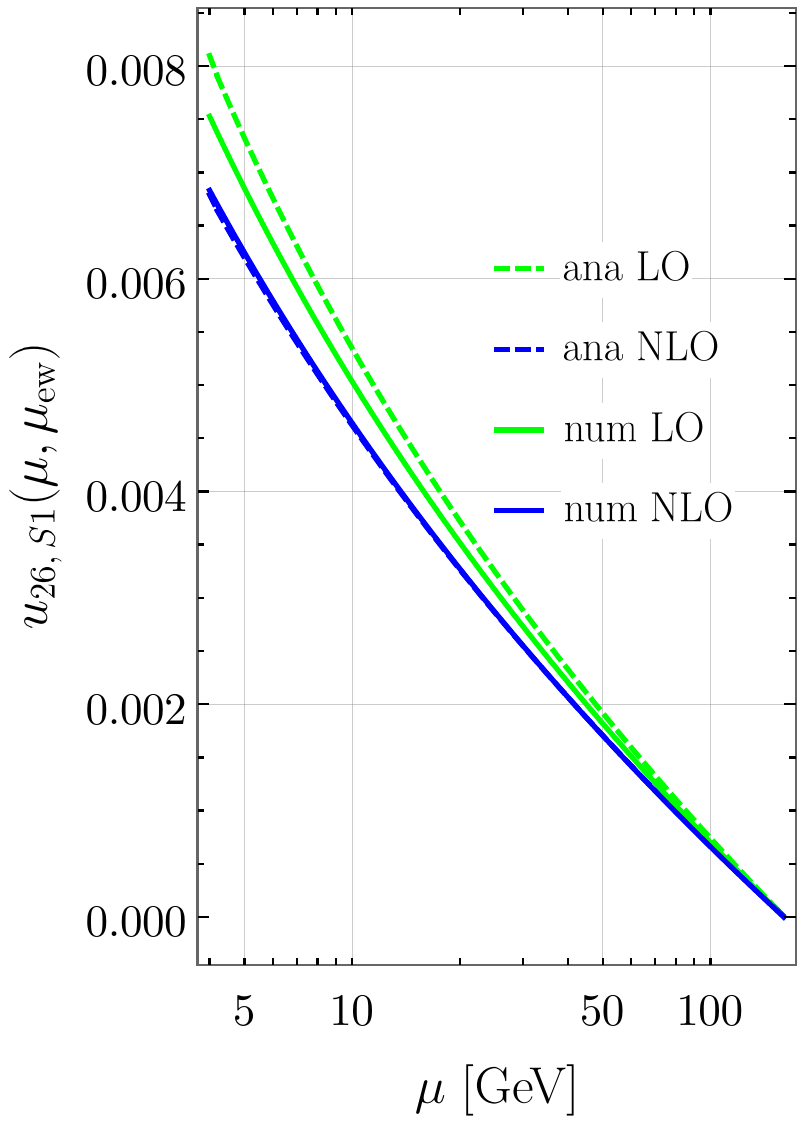}
  \hskip 0.01\textwidth
  \includegraphics[width=0.235\textwidth]{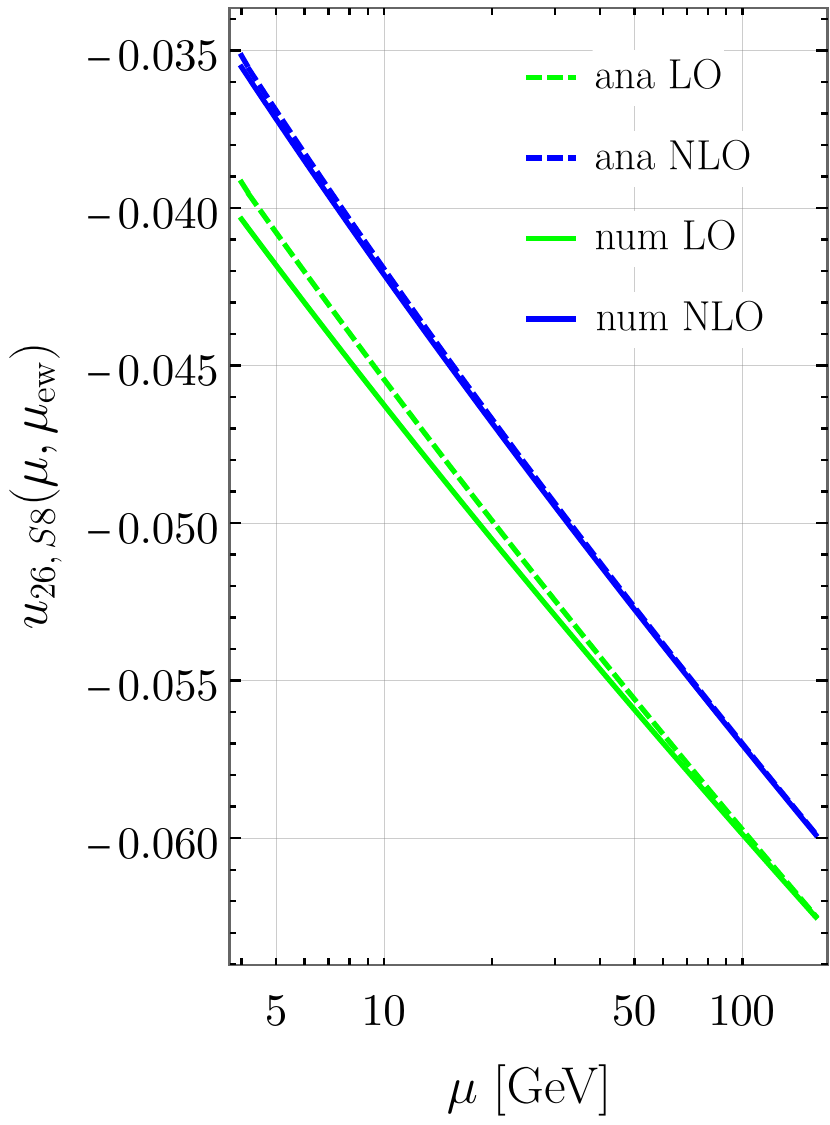}
\caption{\small
  \label{fig:u_ab-SRR,i} The $\mu$-dependence of the quantities
  $u_{ab}(\mu, \muEW)$ at the LO [green] and NLO [blue] for the
  {SRL$,u$ [upper row] and the SRR$,i$ [lower row] sub-sectors,}
  in the $\Nf = 5$ theory. Numerical and analytical solutions of
  the RG equations are shown with the solid and dashed lines,
  respectively.
}
\end{figure}

As examples, we inspect closer the SRL$,u$, the SRR$,i$ and
the SRR,$u$ sub-sectors given in {Eqs.~\eqref{eq:def-BMU-SRL-ops},
\eqref{eq:BMU-SRR,i} and \eqref{eq:def-BMU-SRR-ops}, respectively}.
{They consist of the following} operators:
{\begin{align*}
  & \text{SRL},u :
&
  \big\{ \opL[V1,RL]{uddu}{1ji1}^\dagger,\; \opL[V8,RL]{uddu}{1ji1}^\dagger \big\} &
&
  \leftrightarrow &
&
  & \big\{ \OpL{19}, \; \OpL{20} \big\} ,
\\
  & \text{SRR},i :
&
  \big\{ \opL[S1,RR]{dd}{jiii},\; \opL[S8,RR]{dd}{jiii} \big\} &
&
  \leftrightarrow &
&
  & \big\{ \OpL{25}, \; \OpL{26} \big\} ,
\\
  & \text{SRR},u :
&
  \big\{ \opL[S1,RR]{ud}{11ji},\; \opL[S8,RR]{ud}{11ji},\;
         \opL[S8,RR]{uddu}{1ij1},\; \opL[S8,RR]{uddu}{1ij1} \big\} &
&
  \leftrightarrow &
&
  & \big\{ \OpL{29}, \; \OpL{30}, \; \OpL{31}, \; \OpL{32} \big\} ,
\end{align*}}
in the JMS and BMU bases, respectively. {We note that the results 
for the sectors SRL$,Q$ and SRR$,Q$ with $Q = c ,d_{k\neq i,j}$ are
identical to the results of the SRL$,u$ and SRR$,u$ sectors, respectively.
Furthermore, the results for the SRR$,i$ sector hold for the SRR$,j$ sector, too.}
Our scale choices in the numerical examples are $\muEW = 160 \GeV$ and
$\muLow = 4.2 \GeV$.

{First, we investigate the impact of the NLO corrections on the transformation
matrix  $\hat M_\text{JMS}$ between the JMS and BMU bases, as defined in
Eq.~\eqref{eq:wc-trafo}. We find at $\muEW$} numerically
\begingroup
\setlength\arraycolsep{0.2cm}
{\begin{align}
  & 
& &
  \hat M_\text{JMS,LO}(\muEW)
& &
  \hat M_\text{JMS,NLO}(\muEW)
\notag
\\
  & \text{SRL},u:
& &
  \begin{pmatrix}
   -2 & \frac{1}{3} \\[0.15cm]
    0 & -1
  \end{pmatrix} ,
& &
  \begin{pmatrix}
   -2 & \frac{1}{3} \\[0.15cm]
    0 & -1
  \end{pmatrix} ,
\\
  & \text{SRR},i:
& &
  \begin{pmatrix}
   1  & -0.416\bar{6} \\[0.15cm]
   0  & -0.0625
  \end{pmatrix} ,
& &
  \begin{pmatrix}
   1  & -0.4300 \\[0.15cm]
   0  & -0.0599
  \end{pmatrix} ,
\\
  & \text{SRR},u:
& &
  \begin{pmatrix}
   0 &  \frac{1}{2} & -0.5   &  0.08\bar{3} \\[0.15cm]
   1 & -\frac{1}{6} & 0      & -0.250       \\[0.15cm]
   0 & 0            & -0.125 &  0.0208      \\[0.15cm]
   0 & 0            & 0      & -0.0625
  \end{pmatrix} ,
& & 
  \begin{pmatrix}
   0 &  \frac{1}{2} & -0.505 &  0.070  \\[0.15cm]
   1 & -\frac{1}{6} & -0.037 & -0.259  \\[0.15cm]
   0 & 0            & -0.124 &  0.0215 \\[0.15cm]
   0 & 0            &  0.002 & -0.0623
  \end{pmatrix} ,
\end{align}}
\endgroup
at the LO and NLO in QCD, respectively. {No evanescent operators are
involved in the transformation of the SRL,$u$ sector, which makes the corresponding
NLO corrections vanish.\footnote{{This sector corresponds to the block
$0_{16\times6}$ in Eq.~\eqref{eq:def-mat-delX}, together with the sectors SRL$,c$ 
and SRL$,d_{k\neq i,j}$.}}} The NLO effects {in the SRR$,i$ sector} amount
to a relative change of $+3\%$ in the~(1,2) element and $-4\%$ in the (2,2)
element. {In the SRR$,u$ sector, the NLO corrections give rise to two
non-vanishing entries (2,3) and (4,3). The largest modifications are
in the elements (1,4) and (2,4), amounting to $-16\%$ and $+4\%$, respectively.
%
}
The corresponding $u_{ab}(\muLow,\muEW)$ are the ones with 
{\begin{align*}
  \text{SRL},u: & 
&  a & = \{19, 20\}, & 
   b & = \{V1, V8\},
\\
  \text{SRR},i: & 
&  a & = \{25, 26\}, & 
   b & = \{S1, S8\},
\end{align*}
for the SRL$,u$ and SRR$,i$ sectors, respectively.}.\footnote{Since we consider
{particular sub-sectors} here, somewhat abbreviated JMS operator
names are being used.} Their $\mu$-dependence is shown in
\reffig{fig:u_ab-SRR,i}.  Besides the numerical solution, also the
analytical solution is shown as dashed lines. The difference between
the numerical and analytical solutions is larger at the LO, and
becomes reduced at the NLO, {which indicates reduction of
renormalization-scheme dependences. The numerical effects we observe
correspond to the particular renormalization scheme we have chosen
for the Wilson coefficients and the running couplings. In particular,
we always use the three-loop RG equations for the running of $\alS$,
irrespectively of whether the RG evolution of the Wilson coefficients
is performed at the LO or NLO. Otherwise the shifts from the LO to the NLO
would be much larger but unrelated to the corrections we have calculated.
Let us note that differences between the numerical and analytical solutions of
the NLO RG equations for the Wilson coefficients} are formally of higher
order but should not be misinterpreted as estimates of the overall higher-order
effects. 

The plots remind us impressively that the RG evolution is an important
effect which leads to sizable changes of various Wilson coefficients
when going from $\muEW$ to $\muLow$. The impact of NLO corrections for
the numerical solution is summarized in \reftab{tab:u_ab-SRR,i} for
$\muLow$. We observe corrections in the ballpark {of $10\%$ to
$25\%$} w.r.t.\ the LO result {for our choice of the
renormalization scheme}. The complete list of $u_{ab}(\muLow, \muEW)$
at the NLO for our choice of $\muEW$ and $\muLow$ is given in
\reftab{tab:u_ab-nf-5}. For comparison, we provide the LO results {in
\reftab{tab:u_ab-nf-5} in parenthesis.} {One can see
there that the NLO corrections in the SRR$,Q$ sectors are even larger,
being in the ballpark of $10\%$ to $70\%$, with an extreme case of $>
200\%$ for one element.  The size of NLO corrections in the
VLL$\times$VLL and VLL$\times$VLR blocks range from a few percent
up to $> 100\%$ for some entries. Large NLO corrections are observed
for $u_{ab}$ entries whose absolute values at the LO are small. In the
VLR$\times$VLR block, the NLO corrections do not exceed $40\%$. }

We note that inclusion of the QED and QED$\times$QCD ADMs in
the RG evolution modifies the existing entries {in \reftab{tab:u_ab-nf-5}}
at the level of $1\%$ or less, with a few exceptions up to $3\%$.
Furthermore, there are additional non-vanishing entries for the QED penguin
operators $\OpL{7,8,9,10}$. They are all very small compared
to the ones generated by the QCD mixing. As mentioned before, their
numerical relevance strongly depends on both the observable and
the BSM scenario under consideration.

\begin{table}
\centering
\renewcommand{\arraystretch}{1.3}
\begin{tabular}{c|cccc|cccc}
\hline\hline
& \multicolumn{4}{c|}{SRL$,u$}
& \multicolumn{4}{c}{SRR$,i$}
\\
& $u_{19, V1}$
& $u_{20, V1}$
& $u_{19, V8}$
& $u_{20, V8}$
& $u_{25, S1}$
& $u_{26, S1}$
& $u_{25, S8}$
& $u_{26, S8}$
\\
\hline
& \multicolumn{8}{c}{$\Nf = 5$, $\muLow = 4.2\GeV$}
\\
  LO
& $-1.831$
& $-0.738$
& $0.305$
& $-1.900$
&  1.55
&  0.00737
& $-0.533$
& $-0.0407$
\\
  NLO
& $-1.842$
& $-0.886$
& $0.227$
& $-2.107$
&  1.65
&  0.00670
& $-0.592$
& $-0.0359$
\\
rel.
& $< 1\%$
& $+20\%$
& $-25\%$
& $+11\%$
& $+ 7\%$
& $- 9\%$
& $+11\%$
& $-12\%$
\\
\hline
& \multicolumn{8}{c}{{$\Nf = 3$, $\muLow = 1.3\GeV$}}
\\
  LO
& $-1.740$
& $-1.449$
&  $0.290$
& $-2.801$
&  $1.984$
&  $0.012$
& $-0.646$
& $-0.033$
\\
  NLO
& $-1.815$
& $-1.984$
&  $0.052$
& $-3.531$
&  $2.302$
&  $0.010$
& $-0.779$
& $-0.026$
\\
rel.
& $ +4\%$
& $+37\%$
& $-80\%$
& $+26\%$
& $+16\%$
& $-17\%$
& $+21\%$
& $-21\%$
\\
\hline\hline
\end{tabular}
\renewcommand{\arraystretch}{1.0}
\caption{\small
  \label{tab:u_ab-SRR,i} The impact of NLO QCD corrections on
  $u_{ab}(\mu, \muEW)$ for the SRL$,u$ and SRR$,i$ sub-sectors
  in the $\Nf = 5$ theory for $\muLow = 4.2 \GeV$ and 
  $\Nf = 3$ theory for $\muLow = 1.3 \GeV$. The relative impact is
  indicated w.r.t.\ to $|u_{ab}|$, i.e. $\left(|u_{ab}^\text{NLO}|/|u_{ab}^\text{LO}|-1\right)$
  is displayed in the rows labelled by ``rel.''.
}
\end{table}

\begin{table}
\begin{adjustbox}{width=1\textwidth}
\begin{tabular}{c rr rr rr rr}
\toprule
$C_i(\mu_5)$ & $[C_{ud}^{V1,LL}]_{11ji}$ & $[C_{ud}^{V8,LL}]_{11ji}$ & $[C_{ud}^{V1,LL}]_{22ji}$ & $[C_{ud}^{V8,LL}]_{22ji}$ & $[C_{dd}^{VLL}]_{jikk}$ & $[C_{dd}^{VLL}]_{jkki}$ & $[C_{dd}^{VLL}]_{jiii}$ & $[C_{dd}^{VLL}]_{jijj}$ \\
\midrule
    $C_1$    &            1.139 ( 1.131) &           -0.340 (-0.335) &           -1.139 (-1.131) &            0.340 ( 0.335) &                      -  &                      -  &                      -  &                      -  \\
    $C_2$    &           -0.299 (-0.293) &            0.620 ( 0.614) &            0.299 ( 0.293) &           -0.620 (-0.614) &                      -  &                      -  &                      -  &                      -  \\
    $C_3$    &           -0.007 (-0.002) &            0.010 ( 0.007) &            0.374 ( 0.375) &           -0.106 (-0.105) &          0.753 ( 0.752) &         -0.188 (-0.182) &          0.009 ( 0.012) &          0.009 ( 0.012) \\
    $C_4$    &            0.008 ( 0.004) &           -0.022 (-0.016) &           -0.096 (-0.094) &            0.191 ( 0.189) &         -0.195 (-0.191) &          0.737 ( 0.724) &         -0.033 (-0.026) &         -0.033 (-0.026) \\
    $C_9$    &                        -  &                        -  &            0.760 ( 0.754) &           -0.226 (-0.223) &         -0.760 (-0.754) &          0.199 ( 0.195) &                      -  &                      -  \\
    $C_{10}$ &                        -  &                        -  &           -0.199 (-0.195) &            0.413 ( 0.410) &          0.199 ( 0.195) &         -0.760 (-0.754) &                      -  &                      -  \\
    $C_{11}$ &                        -  &                        -  &                        -  &                        -  &         -0.840 (-0.838) &         -0.840 (-0.838) &          0.420 ( 0.419) &          0.420 ( 0.419) \\
    $C_{14}$ &                        -  &                        -  &                        -  &                        -  &                      -  &                      -  &          0.420 ( 0.419) &         -0.420 (-0.419) \\
    $C_5$    &            0.003 (-0.001) &            0.004 ( 0.004) &            0.004 (-0.001) &            0.002 ( 0.004) &          0.004 (-0.001) &          0.004 ( 0.009) &          0.012 ( 0.007) &          0.012 ( 0.007) \\
    $C_6$    &            0.009 ( 0.005) &           -0.028 (-0.020) &            0.005 ( 0.005) &           -0.020 (-0.020) &          0.006 ( 0.005) &         -0.034 (-0.038) &         -0.040 (-0.033) &         -0.040 (-0.033) \\
\bottomrule
\end{tabular}
\end{adjustbox}

\begin{adjustbox}{width=1\textwidth}
\begin{tabular}{c rr rr rr rr}
\toprule
$C_i(\mu_5)$ & $[C_{du}^{V1,LR}]_{ji11}$ & $[C_{du}^{V8,LR}]_{ji11}$ & $[C_{du}^{V1,LR}]_{ji22}$ & $[C_{du}^{V8,LR}]_{ji22}$ & $[C_{dd}^{V1,LR}]_{jikk}$ & $[C_{dd}^{V8,LR}]_{jikk}$ & $[C_{dd}^{V1,LR}]_{jiii}$ & $[C_{dd}^{V8,LR}]_{jiii}$ \\
\midrule
    $C_3$    &            0.006 ( 0.002) &            0.010 ( 0.009) &            0.006 ( 0.002) &            0.010 ( 0.009) &            0.006 ( 0.002) &            0.010 ( 0.009) &            0.006 ( 0.002) &            0.010 ( 0.009) \\
    $C_4$    &           -0.004 (-0.005) &           -0.021 (-0.020) &           -0.004 (-0.005) &           -0.021 (-0.020) &           -0.004 (-0.005) &           -0.021 (-0.020) &           -0.004 (-0.005) &           -0.021 (-0.020) \\
    $C_5$    &            0.150 ( 0.154) &           -0.014 (-0.020) &            0.150 ( 0.154) &           -0.014 (-0.020) &            0.610 ( 0.612) &           -0.071 (-0.096) &           -0.003 ( 0.001) &            0.005 ( 0.006) \\
    $C_6$    &            0.068 ( 0.056) &            0.144 ( 0.134) &            0.068 ( 0.056) &            0.144 ( 0.134) &            0.290 ( 0.242) &            0.672 ( 0.610) &           -0.006 (-0.006) &           -0.032 (-0.025) \\
    $C_7$    &            0.307 ( 0.305) &           -0.038 (-0.051) &            0.307 ( 0.305) &           -0.038 (-0.051) &           -0.614 (-0.610) &            0.075 ( 0.102) &                        -  &                        -  \\
    $C_8$    &            0.148 ( 0.124) &            0.352 ( 0.317) &            0.148 ( 0.124) &            0.352 ( 0.317) &           -0.297 (-0.247) &           -0.705 (-0.635) &                        -  &                        -  \\
    $C_{12}$ &                        -  &                        -  &                        -  &                        -  &           -0.445 (-0.371) &           -1.057 (-0.952) &            0.223 ( 0.185) &            0.528 ( 0.476) \\
    $C_{13}$ &                        -  &                        -  &                        -  &                        -  &           -0.921 (-0.915) &            0.113 ( 0.153) &            0.460 ( 0.458) &           -0.057 (-0.076) \\
    $C_{15}$ &                        -  &                        -  &                        -  &                        -  &                        -  &                        -  &            0.223 ( 0.185) &            0.528 ( 0.476) \\
    $C_{16}$ &                        -  &                        -  &                        -  &                        -  &                        -  &                        -  &            0.460 ( 0.458) &           -0.057 (-0.076) \\
    $C_{17}$ &            0.223 ( 0.185) &            0.528 ( 0.476) &           -0.223 (-0.185) &           -0.528 (-0.476) &                        -  &                        -  &                        -  &                        -  \\
    $C_{18}$ &            0.460 ( 0.458) &           -0.057 (-0.076) &           -0.460 (-0.458) &            0.057 ( 0.076) &                        -  &                        -  &                        -  &                        -  \\
\bottomrule
\end{tabular}
\end{adjustbox}

\begin{adjustbox}{width=1\textwidth}
\begin{tabular}{c rr rr rr rr}
\toprule
$C_i(\mu_5)$ & $[C_{dd}^{V1,LR}]_{jijj}$ & $[C_{dd}^{V8,LR}]_{jijj}$ & $[C_{uddu}^{V1,LR}]_{1ji1}$ & $[C_{uddu}^{V8,LR}]_{1ji1}$ & $[C_{uddu}^{V1,LR}]_{2ji2}$ & $[C_{uddu}^{V8,LR}]_{2ji2}$ & $[C_{dd}^{V1,LR}]_{jkki}$ & $[C_{dd}^{V8,LR}]_{jkki}$ \\
\midrule
    $C_3$    &            0.006 ( 0.002) &            0.010 ( 0.009) &                          -  &                          -  &                          -  &                          -  &                        -  &                        -  \\
    $C_4$    &           -0.004 (-0.005) &           -0.021 (-0.020) &                          -  &                          -  &                          -  &                          -  &                        -  &                        -  \\
    $C_5$    &           -0.003 ( 0.001) &            0.005 ( 0.006) &                          -  &                          -  &                          -  &                          -  &                        -  &                        -  \\
    $C_6$    &           -0.006 (-0.006) &           -0.032 (-0.025) &                          -  &                          -  &                          -  &                          -  &                        -  &                        -  \\
    $C_{12}$ &            0.223 ( 0.185) &            0.528 ( 0.476) &                          -  &                          -  &                          -  &                          -  &                        -  &                        -  \\
    $C_{13}$ &            0.460 ( 0.458) &           -0.057 (-0.076) &                          -  &                          -  &                          -  &                          -  &                        -  &                        -  \\
    $C_{15}$ &           -0.223 (-0.185) &           -0.528 (-0.476) &                          -  &                          -  &                          -  &                          -  &                        -  &                        -  \\
    $C_{16}$ &           -0.460 (-0.458) &            0.057 ( 0.076) &                          -  &                          -  &                          -  &                          -  &                        -  &                        -  \\
    $C_{19}$ &                        -  &                        -  &             -1.841 (-1.831) &              0.226 ( 0.305) &                          -  &                          -  &                        -  &                        -  \\
    $C_{20}$ &                        -  &                        -  &             -0.890 (-0.741) &             -2.114 (-1.904) &                          -  &                          -  &                        -  &                        -  \\
    $C_{21}$ &                        -  &                        -  &                          -  &                          -  &             -1.841 (-1.831) &              0.226 ( 0.305) &                        -  &                        -  \\
    $C_{22}$ &                        -  &                        -  &                          -  &                          -  &             -0.890 (-0.741) &             -2.114 (-1.904) &                        -  &                        -  \\
    $C_{23}$ &                        -  &                        -  &                          -  &                          -  &                          -  &                          -  &           -1.841 (-1.831) &            0.226 ( 0.305) \\
    $C_{24}$ &                        -  &                        -  &                          -  &                          -  &                          -  &                          -  &           -0.890 (-0.741) &           -2.114 (-1.904) \\
\bottomrule
\end{tabular}
\end{adjustbox}

\begin{adjustbox}{width=1\textwidth}
\begin{tabular}{c rr rr rr rr}
\toprule
$C_i(\mu_5)$ & $[C_{dd}^{S1,RR}]_{jiii}$ & $[C_{dd}^{S8,RR}]_{jiii}$ & $[C_{dd}^{S1,RR}]_{jijj}$ & $[C_{dd}^{S8,RR}]_{jijj}$ & $[C_{ud}^{S1,RR}]_{11ji}$ & $[C_{ud}^{S8,RR}]_{11ji}$ & $[C_{uddu}^{S1,RR}]_{1ij1}$ & $[C_{uddu}^{S8,RR}]_{1ij1}$ \\
\midrule
    $C_{25}$ &             1.656 (1.548) &           -0.593 (-0.534) &                         - &                         - &                         - &                         - &                           - &                           - \\
    $C_{26}$ &             0.007 (0.007) &           -0.036 (-0.041) &                         - &                         - &                         - &                         - &                           - &                           - \\
    $C_{27}$ &                         - &                         - &             1.656 (1.548) &           -0.593 (-0.534) &                         - &                         - &                           - &                           - \\
    $C_{28}$ &                         - &                         - &             0.007 (0.007) &           -0.036 (-0.041) &                         - &                         - &                           - &                           - \\
    $C_{29}$ &                         - &                         - &                         - &                         - &             0.143 ( 0.060)&            0.506 ( 0.495) &             -1.027 (-0.878) &         $<|0.001|$ ( 0.016) \\
    $C_{30}$ &                         - &                         - &                         - &                         - &             2.465 ( 2.089)&           -0.104 (-0.139) &             -0.936 (-0.541) &             -0.464 (-0.395) \\
    $C_{31}$ &                         - &                         - &                         - &                         - &             0.115 ( 0.085)&            0.033 ( 0.025) &             -0.316 (-0.275) &              0.020 ( 0.022) \\
    $C_{32}$ &                         - &                         - &                         - &                         - &            -0.034 (-0.028)&            0.006 ( 0.009) &              0.046 ( 0.035) &             -0.046 (-0.050) \\
\bottomrule
\end{tabular}
\end{adjustbox}

\begin{adjustbox}{width=1\textwidth}
\begin{tabular}{c rr rr rr rr}
\toprule
$C_i(\mu_5)$ & $[C_{ud}^{S1,RR}]_{22ji}$ & $[C_{ud}^{S8,RR}]_{22ji}$ & $[C_{uddu}^{S1,RR}]_{2ij2}$ & $[C_{uddu}^{S8,RR}]_{2ij2}$ & $[C_{dd}^{S1,RR}]_{jikk}$ & $[C_{dd}^{S8,RR}]_{jikk}$ & $[C_{dd}^{S1,RR}]_{jkki}$ & $[C_{dd}^{S8,RR}]_{jkki}$ \\
\midrule
    $C_{33}$ &            0.143 ( 0.060) &            0.506 ( 0.495) &             -1.027 (-0.878) &         $<|0.001|$ ( 0.016) &                         - &                         - &                         - &                         - \\
    $C_{34}$ &            2.465 ( 2.089) &           -0.104 (-0.139) &             -0.936 (-0.541) &             -0.464 (-0.395) &                         - &                         - &                         - &                         - \\
    $C_{35}$ &            0.115 ( 0.085) &            0.033 ( 0.025) &             -0.316 (-0.275) &              0.020 ( 0.022) &                         - &                         - &                         - &                         - \\
    $C_{36}$ &           -0.034 (-0.028) &            0.006 ( 0.009) &              0.046 ( 0.035) &             -0.046 (-0.050) &                         - &                         - &                         - &                         - \\
    $C_{37}$ &                         - &                         - &                           - &                           - &            0.143 ( 0.060) &            0.506 ( 0.495) &           -1.027 (-0.878) &       $<|0.001|$ ( 0.016) \\
    $C_{38}$ &                         - &                         - &                           - &                           - &            2.465 ( 2.089) &           -0.104 (-0.139) &           -0.936 (-0.541) &           -0.464 (-0.395) \\
    $C_{39}$ &                         - &                         - &                           - &                           - &            0.115 ( 0.085) &            0.033 ( 0.025) &           -0.316 (-0.275) &            0.020 ( 0.022) \\
    $C_{40}$ &                         - &                         - &                           - &                           - &           -0.034 (-0.028) &            0.006 ( 0.009) &            0.046 ( 0.035) &           -0.046 (-0.050) \\
\bottomrule
\end{tabular}
\end{adjustbox}

\caption{\label{tab:u_ab-nf-5}
  The elements $u_{ab}(\muLow,\muEW)$ of the NLO (LO) RG evolution operator in
  Eq.~\eqref{eq:def-u-nf5} for $\Nf = 5$. They connect the JMS Wilson
  coefficients at $\muEW = 160 \GeV$ to the BMU ones at $\muLow =
  \mu_5 = 4.2 \GeV$.
}
\end{table}

\begin{table}[H]
\begin{adjustbox}{width=1\textwidth}\begin{tabular}{c rr rr rr rr}
\toprule
$C_i(\mu_4)$ & $[C_{ud}^{V1,LL}]_{11ji}$ & $[C_{ud}^{V8,LL}]_{11ji}$ & $[C_{ud}^{V1,LL}]_{22ji}$ & $[C_{ud}^{V8,LL}]_{22ji}$ & $[C_{dd}^{VLL}]_{jikk}$ & $[C_{dd}^{VLL}]_{jkki}$ & $[C_{dd}^{VLL}]_{jiii}$ & $[C_{dd}^{VLL}]_{jijj}$ \\
\midrule
    $C_1$    &            1.279 ( 1.258) &           -0.474 (-0.462) &           -1.279 (-1.258) &            0.474 ( 0.462) &                      -  &                      -  &                      -  &                      -  \\
    $C_2$    &           -0.522 (-0.505) &            0.727 ( 0.713) &            0.522 ( 0.505) &           -0.727 (-0.713) &                      -  &                      -  &                      -  &                      -  \\
    $C_3$    &           -0.018 (-0.005) &            0.020 ( 0.014) &            0.408 ( 0.414) &           -0.138 (-0.140) &          0.242 ( 0.240) &         -0.935 (-0.916) &          0.316 ( 0.319) &          0.316 ( 0.319) \\
    $C_4$    &            0.021 ( 0.011) &           -0.038 (-0.027) &           -0.153 (-0.157) &            0.204 ( 0.211) &         -0.246 (-0.238) &          0.933 ( 0.905) &         -0.098 (-0.087) &         -0.098 (-0.087) \\
    $C_9$    &                        -  &                        -  &            0.853 ( 0.839) &           -0.316 (-0.308) &         -0.250 (-0.242) &          0.951 ( 0.934) &         -0.302 (-0.298) &         -0.302 (-0.298) \\
    $C_{10}$ &                        -  &                        -  &           -0.348 (-0.337) &            0.484 ( 0.476) &          0.250 ( 0.242) &         -0.951 (-0.934) &          0.049 ( 0.047) &          0.049 ( 0.047) \\
    $C_{14}$ &                        -  &                        -  &                        -  &                        -  &                      -  &                      -  &          0.379 ( 0.377) &         -0.379 (-0.377) \\
    $C_5$    &            0.006 (-0.003) &            0.004 ( 0.007) &            0.005 (-0.003) &            0.004 ( 0.007) &      0.005 ($<|0.001|$) &         -0.003 ( 0.007) &          0.015 ( 0.010) &          0.015 ( 0.010) \\
    $C_6$    &            0.028 ( 0.014) &           -0.059 (-0.040) &            0.027 ( 0.014) &           -0.057 (-0.040) &          0.012 ( 0.006) &         -0.056 (-0.051) &         -0.077 (-0.061) &         -0.077 (-0.061) \\
\bottomrule
\end{tabular}
\end{adjustbox}

\begin{adjustbox}{width=1\textwidth}\begin{tabular}{c rr rr rr rr}
\toprule
$C_i(\mu_4)$ & $[C_{du}^{V1,LR}]_{ji11}$ & $[C_{du}^{V8,LR}]_{ji11}$ & $[C_{du}^{V1,LR}]_{ji22}$ & $[C_{du}^{V8,LR}]_{ji22}$ & $[C_{dd}^{V1,LR}]_{jikk}$ & $[C_{dd}^{V8,LR}]_{jikk}$ & $[C_{dd}^{V1,LR}]_{jiii}$ & $[C_{dd}^{V8,LR}]_{jiii}$ \\
\midrule
    $C_3$    &            0.017 ( 0.006) &            0.023 ( 0.020) &            0.022 ( 0.006) &            0.034 ( 0.020) &            0.009 ( 0.003) &            0.016 ( 0.011) &            0.017 ( 0.006) &            0.023 ( 0.020) \\
    $C_4$    &           -0.015 (-0.014) &           -0.043 (-0.040) &           -0.033 (-0.014) &           -0.074 (-0.040) &           -0.009 (-0.005) &           -0.028 (-0.019) &           -0.015 (-0.014) &           -0.043 (-0.040) \\
    $C_5$    &            0.296 ( 0.293) &           -0.003 (-0.037) &       $<|0.001|$ ( 0.004) &            0.015 ( 0.011) &           -0.003 ( 0.001) &            0.002 ( 0.004) &            0.296 ( 0.293) &           -0.003 (-0.037) \\
    $C_6$    &            0.312 ( 0.231) &            0.518 ( 0.420) &           -0.045 (-0.018) &           -0.116 (-0.057) &           -0.016 (-0.007) &           -0.061 (-0.033) &            0.312 ( 0.231) &            0.518 ( 0.420) \\
    $C_7$    &            0.605 ( 0.579) &           -0.015 (-0.096) &                         - &                         - &                         - &                         - &           -0.303 (-0.289) &            0.008 ( 0.048) \\
    $C_8$    &            0.678 ( 0.498) &            1.205 ( 0.953) &                         - &                         - &                         - &                         - &           -0.339 (-0.249) &           -0.602 (-0.477) \\
    $C_{15}$ &                         - &                         - &                         - &                         - &                         - &                         - &            0.509 ( 0.374) &            0.904 ( 0.715) \\
    $C_{16}$ &                         - &                         - &                         - &                         - &                         - &                         - &            0.454 ( 0.434) &           -0.012 (-0.072) \\
\bottomrule
\end{tabular}
\end{adjustbox}

\begin{adjustbox}{width=1\textwidth}\begin{tabular}{c rr rr rr rr}
\toprule
$C_i(\mu_4)$ & $[C_{dd}^{V1,LR}]_{jijj}$ & $[C_{dd}^{V8,LR}]_{jijj}$ & $[C_{uddu}^{V1,LR}]_{1ji1}$ & $[C_{uddu}^{V8,LR}]_{1ji1}$ & $[C_{uddu}^{V1,LR}]_{2ji2}$ & $[C_{uddu}^{V8,LR}]_{2ji2}$ & $[C_{dd}^{V1,LR}]_{jkki}$ & $[C_{dd}^{V8,LR}]_{jkki}$ \\
\midrule
    $C_3$    &            0.017 ( 0.006) &            0.023 ( 0.020) &                           - &                           - &                           - &                           - &                         - &                         - \\
    $C_4$    &           -0.015 (-0.014) &           -0.043 (-0.040) &                           - &                           - &                           - &                           - &                         - &                         - \\
    $C_5$    &            0.296 ( 0.293) &           -0.003 (-0.037) &                           - &                           - &                           - &                           - &                         - &                         - \\
    $C_6$    &            0.312 ( 0.231) &            0.518 ( 0.420) &                           - &                           - &                           - &                           - &                         - &                         - \\
    $C_7$    &           -0.303 (-0.289) &            0.008 ( 0.048) &                           - &                           - &                           - &                           - &                         - &                         - \\
    $C_8$    &           -0.339 (-0.249) &           -0.602 (-0.477) &                           - &                           - &                           - &                           - &                         - &                         - \\
    $C_{15}$ &           -0.509 (-0.374) &           -0.904 (-0.715) &                           - &                           - &                           - &                           - &                         - &                         - \\
    $C_{16}$ &           -0.454 (-0.434) &            0.012 ( 0.072) &                           - &                           - &                           - &                           - &                         - &                         - \\
    $C_{19}$ &                         - &                         - &             -1.815 (-1.736) &              0.046 ( 0.289) &                           - &                           - &                         - &                         - \\
    $C_{20}$ &                         - &                         - &             -2.035 (-1.495) &             -3.615 (-2.860) &                           - &                           - &                         - &                         - \\
\bottomrule
\end{tabular}
\end{adjustbox}

\begin{adjustbox}{width=1\textwidth}\begin{tabular}{c rr rr rr rr}
\toprule
$C_i(\mu_4)$ & $[C_{dd}^{S1,RR}]_{jiii}$ & $[C_{dd}^{S8,RR}]_{jiii}$ & $[C_{dd}^{S1,RR}]_{jijj}$ & $[C_{dd}^{S8,RR}]_{jijj}$ & $[C_{ud}^{S1,RR}]_{11ji}$ & $[C_{ud}^{S8,RR}]_{11ji}$ & $[C_{uddu}^{S1,RR}]_{1ij1}$ & $[C_{uddu}^{S8,RR}]_{1ij1}$ \\
\midrule
    $C_{25}$ &             2.331 (2.010) &           -0.789 (-0.653) &                         - &                         - &                         - &                         - &                           - &                           - \\
    $C_{26}$ &             0.011 (0.012) &           -0.026 (-0.032) &                         - &                         - &                         - &                         - &                           - &                           - \\
    $C_{27}$ &                         - &                         - &             2.331 (2.010) &           -0.789 (-0.653) &                         - &                         - &                           - &                           - \\
    $C_{28}$ &                         - &                         - &             0.011 (0.012) &           -0.026 (-0.032) &                         - &                         - &                           - &                           - \\
    $C_{29}$ &                         - &                         - &                         - &                         - &            0.595 ( 0.205) &            0.593 ( 0.535) &             -1.863 (-1.284) &             -0.047 (-0.014) \\
    $C_{30}$ &                         - &                         - &                         - &                         - &            4.933 ( 3.359) &            0.069 (-0.087) &             -2.904 (-1.349) &             -0.794 (-0.566) \\
    $C_{31}$ &                         - &                         - &                         - &                         - &            0.345 ( 0.208) &            0.077 ( 0.050) &             -0.628 (-0.453) &              0.002 ( 0.016) \\
    $C_{32}$ &                         - &                         - &                         - &                         - &           -0.095 (-0.066) &       $<|0.001|$ ( 0.010) &              0.118 ( 0.078) &             -0.032 (-0.042) \\
\bottomrule
\end{tabular}
\end{adjustbox}
\caption{\label{tab:u_ab-nf-3}
  The elements $u_{ab}(\muLow,\muEW)$ of the NLO (LO) RG evolution operator in
  Eq.~\eqref{eq:def-u-nf3} for $\Nf = 3$ at $\muLow$. They connect the
  JMS Wilson coefficients at $\muEW = 160 \GeV$ to the BMU ones at
  $\muLow = \mu_4 = 1.3 \GeV$.
}
\end{table}

%
%
\subsection[$K\to \pi\pi$ and $\epe$]
{\bm $K\to \pi\pi$ and $\epe$}
\label{sec:epe}

Non-leptonic Kaon decays are especially sensitive to CP-violating effects
beyond the SM because of the strong suppression of the SM contribution.
In particular, the observables $\varepsilon$ and $\varepsilon'$ measure
indirect and direct CP violation in $\KKbar$ mixing and $K^0$ decay into
$\pi\pi$, respectively. The world average from NA48~\cite{Batley:2002gn}
and KTeV~\cite{AlaviHarati:2002ye, Abouzaid:2010ny} collaborations for their
ratio $\epe$ reads
\begin{align}
  \label{EXP}
  (\epe)_\text{exp} & = (16.6\pm 2.3) \times 10^{-4} .
\end{align}

The $K\to\pi\pi$ hadronic matrix elements of the SM operators
contributing to the ratio $\epe$ from RBC-UKQCD Lattice QCD
collaboration~\cite{Abbott:2020hxn} together with refinements in the
estimate of isospin-breaking effects~\cite{Cirigliano:2019cpi,
Buras:2020pjp} and QCD corrections \cite{Cerda-Sevilla:2016yzo,
Cerda-Sevilla:2018hjk, Aebischer:2019mtr} indicate that although the
SM is compatible with the experimental world average, still significant
room for NP is left in this ratio. The most recent analysis
summarizing the present status of $\epe$ in the SM implies
\cite{Buras:2020pjp, Aebischer:2020jto}

\begin{align}
  \label{SM}
  (\epe)_\text{SM}
    = (13.9 \pm 5.2) \times 10^{-4} \,.
\end{align}
Thus, modifications of $\epe$ by NP as large as $10^{-3}$ are presently
not excluded. In fact, as argued on the basis of the large-$\Nc$ QCD in
Ref.~\cite{Buras:2021ane}, the values of $\epe$ in the SM in the ballpark of
$5\times 10^{-4}$ are still possible.

The ratio $\epe$ puts serious constraints on CP-violating phases in BSM
models. Previous analyses of NP effects in $\epe$ were restricted
to the LO RG evolution of the BSM contributions. For
example, the derivation of a master formula for
$\epe$~\cite{Aebischer:2018quc} or the SMEFT anatomy of
$\epe$~\cite{Aebischer:2018csl} demonstrated possible
interplay of $\epe$ with other quark-flavour and collider
observables. Our current work allows us to include also
the NLO QCD corrections.

Below, we consider elements of the RG evolution operator
\begin{align}
  \label{eq:def-u-nf3}
  u_{ab}(\muLow,\muEW) & \; \equiv \; \big[
  \hat U^{(3)}(\muLow, \mu_4) \, \hat M^{(4)}(\mu_4) \;
  \hat U^{(4)}(\mu_4,  \mu_5) \, \hat M^{(5)}(\mu_5) \;
  \hat U^{(5)}(\mu_5,  \muEW) \;
                        \hat M_\text{JMS}(\muEW) \big]_{ab} \, ,
\end{align}
to demonstrate the numerical impact of the NLO corrections. In
Kaon physics, the $b$-quark and $c$-quark thresholds are included in the
RG evolution. At each threshold, the number of operators is reduced, as
described in \refsec{sec:nonleptonic-ops}, from $2 \times 40$ operators
in the $\Nf = 5$ theory (index $b$) to $2 \times 23$ operators in the
$\Nf=3$ theory (index $a$).\footnote{
We remind that we work with a redundant set of the QCD- and
QED-penguin operators.}
A complete list of $u_{ab}(\muLow,\muEW)$ at the NLO for our choice of
$\muEW$ and $\muLow$ is given in \reftab{tab:u_ab-nf-3}. For comparison,
we provide the results at the LO {in \reftab{tab:u_ab-nf-3} in
parenthesis.}

{
The NLO corrections in the $\Nf=3$ flavour theory are, in general, larger than
in the $\Nf=5$ flavour theory, as a consequence of fast RG evolution of $\alS$ 
at scales reaching $\muLow \approx 1 \GeV$. Some of the $u_{ab}$ entries cross zero
during their RG evolution. In certain cases, we observe
suppressed magnitudes at the intermediate scale $\mu_5$ but again
sizable entries at $\muLow$. In other cases, $u_{ab}$ cross zero in the vicinity of
$\muLow$ -- see, for example, $u_{19,V8}$ and $u_{26,S8}$ in \reftab{tab:u_ab-SRR,i}.
Close to such points, the relative impact
of NLO corrections becomes amplified. Thus, the actual impact of the NLO corrections strongly depends
on the considered entries $u_{ab}$. However, as in the case of $\Nf=5$, their relative
impact is usually the largest for $u_{ab}$ with small absolute values at the LO. 

The impact of our results on the CP-violating ratio $\epe$ is
discussed in a separate publication~\cite{Aebischer:2021hws} by most
co-authors of the present article. The previous BSM master
formula~\cite{Aebischer:2018quc,Aebischer:2018csl} is generalized
there to NLO in QCD. It provides $\epe$ in terms of the Wilson
coefficients of the JMS basis at the electroweak scale $\muEW$. 
Depending on the considered JMS Wilson coefficient, the NLO QCD corrections
in the RG evolution lead to an impact on the corresponding coefficient
in $\epe$ in the ballpark of $(20-40)\%$, reaching $60\%$ in certain cases.
The magnitude of such corrections is thus similar or sometimes even larger
than uncertainties of nonperturbative origin in $K\to \pi\pi$ matrix
elements of the nonleptonic operators.}

%
%
%
\section{Conclusions}
\label{sec:conclusion}

In the present paper, we presented all ingredients that are necessary to write
down the amplitude for any $\Delta F=1$ non-leptonic transition in terms of the electroweak-scale WET Wilson
coefficients in the JMS basis. Our expressions
include the NLO QCD RG evolution of the Wilson coefficients
from the electroweak scale down to the low energy scale at which
hadronic matrix elements are being calculated.
As the latter evolution is most conveniently performed in the BMU basis, an
involved NLO transformation from the JMS basis to the BMU one was necessary.
While rather straightforward at the tree-level, it required
a very careful treatment of evanescent operators to ensure that the corresponding
scheme dependence cancels the one of the two-loop ADMs
that govern the NLO QCD RG evolution.

The latter transformation, that we presented in detail in
\refsec{sec:basis-change}, allows us to obtain expressions for physical observables
in terms of the JMS Wilson coefficients at the electroweak scale. The latter coefficients
are related to the SMEFT ones at the NP scale $\muNP$ via
\begin{align}
  \label{eq:SMEFTWEt-2}
  \vec{\mathcal{C}}_\text{JMS}(\muEW) &
  = \hat K \, \vec{\mathcal{C}}_\text{SMEFT}(\muEW)
  = \hat K \, \hat U_\text{SMEFT}(\muEW,\, \muNP) \, \vec{\mathcal{C}}_\text{SMEFT}(\muNP),
\end{align}
where $\hat K$ is the  matching matrix calculated including one-loop
contributions in Ref.~\cite{Dekens:2019ept}, and $\hat U_\text{SMEFT}(\muEW,\muNP)$ is the
SMEFT RG evolution matrix from $\muNP$ down to $\muEW$. Once the SMEFT two-loop ADMs are
calculated in the future, our results will allow to obtain complete NLO expressions
for amplitudes of all low-energy non-leptonic decays in any BSM scenario where the relevant
new particles have masses of order $\muNP$ or larger.

One should again emphasize that only a combination of the NLO
QCD RG evolution in WET with the one-loop QCD corrections in the
matching of SMEFT onto WET, with the NLO QCD RG evolution
within the SMEFT, and the corresponding one-loop
matching of a BSM scenario onto SMEFT will enable a proper
cancellation of RS dependences that appear at various places in
a complete analysis.

In the present paper, we have restricted ourselves to illustrating
the impact of NLO QCD effects on the WET Wilson coefficients that
matter for non-leptonic Kaon and $B$-meson decays. As a next step, in
a separate paper, an NLO-upgraded master formula for the BSM
contributions to $\epe$ is going to be presented. The upgrade will
improve the one presented in Ref.~\cite{Aebischer:2018quc}.

%
%
%

\section*{Acknowledgements}

J. A. acknowledges financial support from the Swiss National Science Foundation
(Project No. P400P2\_183838).
The work of C.B. is supported by DFG under grant BO-4535/1-1.
A.J.B acknowledges financial support from the Excellence Cluster ORIGINS,
funded by the Deutsche Forschungsgemeinschaft (DFG, German Research Foundation)
under Germany's Excellence Strategy – EXC-2094 – 390783311.
M.M has been partially supported by the National Science Center, Poland, under
the research projects 2017/25/B/ST2/00191, 2020/37/B/ST2/02746,
and the HARMONIA project under contract 2015/18/M/ST2/00518.
J.K. is financially supported by the Alexander von Humboldt Foundation’s postdoctoral
research fellowship.

%
%
%

\appendix

%
%
%
\section{\bm Non-leptonic $\Delta F = 1$ operators}
\label{app:non-leptonic-op's}

In this appendix, we list the baryon-number conserving four-quark $\DF=1$
operators in the JMS and BMU bases. We focus on $\Delta C = 0$ processes
$d_i \to d_j \, Q\bar Q$ with $Q = d_i, d_j, d_{k\neq i, j}, u, c$,
and neglect chromo-magnetic dipole operators.

%
%
%
\subsection{JMS basis}
\label{app:def-JMS}

In the JMS basis~\cite{Jenkins:2017jig}, the full set of
dimension-six four-quark operators is divided into four
classes with different chirality structures: $(\oL{L}L)(\oL{L}L)$,
$(\oL{R}R)(\oL{R}R)$, $(\oL{L}L)(\oL{R}R)$ and $(\oL{L}R)(\oL{L}R)$.
The non-leptonic operators that contain at least one pair of down-type
quarks are given as
\begin{align}
  \notag
  & (\oL{L}L)(\oL{L}L) &
\\ \notag
  \opL[VLL]{dd}{prst} &
  = (\bar{d}_L^p \gamma_\mu d_L^r) (\bar{d}_L^s \gamma^\mu d_L^t) , &
\\
  \opL[V1,LL]{ud}{prst} &
  = (\bar{u}_L^p \gamma_\mu u_L^r) (\bar{d}_L^s \gamma^\mu d_L^t) , &
  \opL[V8,LL]{ud}{prst} &
  = (\bar{u}_L^p \gamma_\mu T^A u_L^r) (\bar{d}_L^s \gamma^\mu T^A d_L^t) ,
\\[0.2cm]
  \notag
  & (\oL{R}R)(\oL{R}R) &
\\
  \notag
  \opL[VRR]{dd}{prst} &
  = (\bar{d}_R^p \gamma_\mu d_R^r) (\bar{d}_R^s \gamma^\mu d_R^t) , &
\\
  \opL[V1,RR]{ud}{prst} &
  = (\bar{u}_R^p \gamma_\mu u_R^r) (\bar{d}_R^s \gamma^\mu d_R^t) , &
  \opL[V8,RR]{ud}{prst} &
  = (\bar{u}_R^p \gamma_\mu T^A u_R^r) (\bar{d}_R^s \gamma^\mu T^A d_R^t) ,
\\[0.2cm]
  \notag
  & (\oL{L}L)(\oL{R}R) &
\\
  \notag
  \opL[V1,LR]{dd}{prst} &
  = (\bar{d}_L^p \gamma_\mu d_L^r) (\bar{d}_R^s \gamma^\mu d_R^t) , &
  \opL[V8,LR]{dd}{prst} &
  = (\bar{d}_L^p \gamma_\mu T^A d_L^r) (\bar{d}_R^s \gamma^\mu T^A d_R^t) ,
\\
  \notag
  \opL[V1,LR]{ud}{prst} &
  = (\bar{u}_L^p \gamma_\mu u_L^r) (\bar{d}_R^s \gamma^\mu d_R^t) , &
  \opL[V8,LR]{ud}{prst} &
  = (\bar{u}_L^p \gamma_\mu T^A u_L^r) (\bar{d}_R^s \gamma^\mu T^A d_R^t) ,
\\
  \notag
  \opL[V1,LR]{du}{prst} &
  = (\bar{d}_L^p \gamma_\mu d_L^r) (\bar{u}_R^s \gamma^\mu u_R^t) , &
  \opL[V8,LR]{du}{prst} &
  = (\bar{d}_L^p \gamma_\mu T^A d_L^r) (\bar{u}_R^s \gamma^\mu T^A u_R^t) ,
\\
  \opL[V1,LR]{uddu}{prst} &
  = (\bar{u}_L^p \gamma_\mu d_L^r) (\bar{d}_R^s \gamma^\mu u_R^t) , &
  \opL[V8,LR]{uddu}{prst} &
  = (\bar{u}_L^p \gamma_\mu T^A d_L^r) (\bar{d}_R^s \gamma^\mu T^A u_R^t) ,
\\[0.2cm]
  \notag
  & (\oL{L}R)(\oL{L}R) &
\\
  \notag
  \opL[S1,RR]{dd}{prst} &
  = (\bar{d}_L^p d_R^r) (\bar{d}_L^s d_R^t) , &
  \opL[S8,RR]{dd}{prst} &
  = (\bar{d}_L^p T^A d_R^r) (\bar{d}_L^s T^A d_R^t) ,
\\
  \notag
  \opL[S1,RR]{ud}{prst} &
  = (\bar{u}_L^p u_R^r) (\bar{d}_L^s d_R^t) , &
  \opL[S8,RR]{ud}{prst} &
  = (\bar{u}_L^p T^A u_R^r) (\bar{d}_L^s T^A d_R^t) ,
\\
  \opL[S1,RR]{uddu}{prst} &
  = (\bar{u}_L^p d_R^r) (\bar{d}_L^s u_R^t) , &
  \opL[S8,RR]{uddu}{prst} &
  = (\bar{u}_L^p T^A d_R^r) (\bar{d}_L^s T^A u_R^t) ,
\end{align}
where $p,r,s,t$ denote quark-flavour indices in the
mass-eigenstate basis. Colour structures are expressed
in terms of the generators $T^A$ of $\SUthreeC$.  The listed
operators include both the ``non-flipped'' and the chirality-flipped
ones.

Some of the above operators do not equal their hermitian conjugates with
permuted generation indices. Although we do not list them explicitly here,
they have to be treated as independent operators. Note that also
$\OpL[V1,LR]{uddu}$ and $\OpL[V8,LR]{uddu}$ require such hermitian conjugates. The
same holds for the operators $(\oL{L}R)(\oL{L}R)$. The JMS basis
contains no operators with $\bar\psi \sigma^{\mu\nu} \psi'$ tensor currents.

%
%
%
\subsection[BMU basis for $\Nf = 5$]
{\bm BMU basis for $\Nf = 5$}
\label{app:def-BMU}

The BMU basis~\cite{Buras:2000if} contains several sub-sectors of operators,
depending on their quark-flavour content and Lorentz structures. They
form separate blocks under RG evolution, thereby minimizing the operator
mixing.

The first 10 elements of the BMU basis form the {\bf SM basis} that consists of two $d_i \to d_j \,u_k \bar u_k$
current-current-type and eight penguin-type operators. The current-current ones read
\begin{equation}
\begin{aligned}
  \OpL{1} &
  = \OpL[\text{VLL},u]{1}
  = (\bar d_j^\alpha \gamma_\mu P_L u^\beta)
    (\bar u^\beta    \gamma^\mu P_L d_i^\alpha) ,
\\
  \OpL{2} &
  = \OpL[\text{VLL},u]{2}
  = (\bar d_j^\alpha \gamma_\mu P_L u^\alpha)
    (\bar u^\beta    \gamma^\mu P_L d_i^\beta) ,
\end{aligned}
\end{equation}
where the naming used in Ref.~\cite{Buras:2000if} is given as
well. The $\SUthreeC$ colour indices are denoted by $\alpha$ and
$\beta$. The choice of $u_k = u$ implies that the corresponding
operators for $u_k = c$ are eliminated in flavour of the QCD-
and QED-penguin operators (see below), which is most suited for
kaon physics. On the other hand, for $B$-meson decays, one
might use instead $u_k = c$, i.e. keep VLL,$c =
(\OpL[\text{VLL},c]{1},\, \OpL[\text{VLL},c]{2})$, and eliminate
the operators with $u_k = u$ in flavour of the penguin operators.

\begin{figure}
\centering
  \begin{subfigure}[t]{0.32\textwidth}
    \centering
    \includegraphics[width=0.7\textwidth]{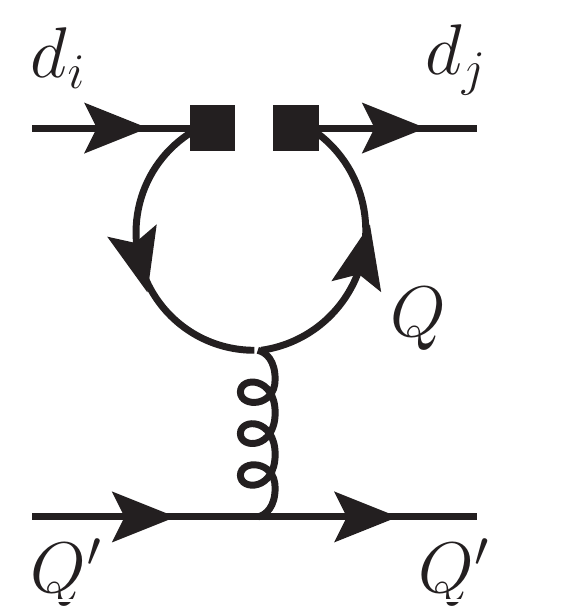}
    \caption{}
    \label{fig:P-ins-open-1l}
  \end{subfigure}
  \begin{subfigure}[t]{0.32\textwidth}
    \centering
    \includegraphics[width=0.7\textwidth]{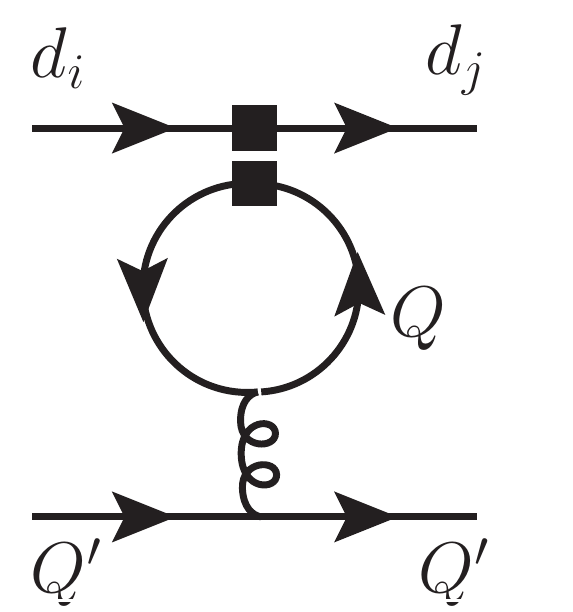}
    \caption{}
    \label{fig:P-ins-closed-1l}
  \end{subfigure}
\caption{\small
  \label{fig:peng-ins-1l}
  The QCD-penguin insertions at one-loop of non-leptonic operators with
  open [left] and closed [right] fermion loops. The inserted operator
  can either be a current-current one (the open loop
  only) or a penguin one (both cases). The analogous
  QED-penguin insertions are given by diagrams where the gluon is
  replaced by a photon.
}
\end{figure}

The penguin-type diagrams (see \reffig{fig:peng-ins-1l})
with insertions of the current-current and penguin
operators can give rise to mixing into the penguin
ones. The eight penguin-type operators with left-handed $(\bar d_j
\gamma_\mu P_L d_i)$ vector currents that are present in
the SM basis can be divided into the QCD-penguin sub-sector
{\bf\bm P$_\text{QCD}$}
\begin{equation}
  \label{eq:QCD-peng-op}
\begin{aligned}
  \OpL{3} & = (\bar d_j^\alpha \gamma_\mu P_L d_i^\alpha)
    \!\sum_{q}(\bar q^\beta    \gamma^\mu P_L \, q^\beta) , \qquad
&
  \OpL{4} & = (\bar d_j^\alpha \gamma_\mu P_L d_i^\beta)
    \!\sum_{q}(\bar q^\beta    \gamma^\mu P_L \, q^\alpha) ,
\\
  \OpL{5} & = (\bar d_j^\alpha \gamma_\mu P_L d_i^\alpha)
    \!\sum_{q}(\bar q^\beta    \gamma^\mu P_R \, q^\beta) ,
&
  \OpL{6} & = (\bar d_j^\alpha \gamma_\mu P_L d_i^\beta)
    \!\sum_{q}(\bar q^\beta    \gamma^\mu P_R \, q^\alpha) ,
\end{aligned}
\end{equation}
and the {\bf\bm P$_\text{QED}$} one
\begin{equation}
  \label{eq:QED-peng-op}
\begin{aligned}
  \OpL{7} & = \frac{3}{2}\,(\bar d_j^\alpha \gamma_\mu P_L d_i^\alpha)
      \!\sum_{q} \! Q_q \, (\bar q^\beta \gamma^\mu P_R \, q^\beta) ,  \qquad
&
  \OpL{8} & = \frac{3}{2}\,(\bar d_j^\alpha \gamma_\mu P_L d_i^\beta)
      \!\sum_{q} \! Q_q \, (\bar q^\beta \gamma^\mu P_R \, q^\alpha) ,
\\
  \OpL{9} & = \frac{3}{2}\,(\bar d_j^\alpha \gamma_\mu P_L d_i^\alpha)
      \!\sum_{q} \! Q_q \, (\bar q^\beta \gamma^\mu P_L \, q^\beta) ,
&
  \OpL{10} & =\frac{3}{2}\,(\bar d_j^\alpha \gamma_\mu P_L d_i^\beta)
      \!\sum_{q} \! Q_q \, (\bar q^\beta \gamma^\mu P_L \, q^\alpha) .
\end{aligned}
\end{equation}
The sums over $q = u,d,s,c,b$ include all the
five active quark flavours whose electromagnetic charges
are denoted by $Q_q$: $Q_u = +2/3$ and $Q_d = -1/3$. Note
that we use here the ``traditional'' convention~\cite{Gilman:1979bc, Buras:1992zv, Ciuchini:1993vr}
for the penguin operators that is suitable up to the NLO in QCD.
Beyond the NLO, the issue of traces with~$\gamma_5$ makes
working with an alternative basis more
convenient~\cite{Chetyrkin:1996vx, Chetyrkin:1997gb}.
Ref.~\cite{Gorbahn:2004my} provides NNLO QCD transformations between
the two conventions for the VLL$,c$ and P$_\text{QCD}$ Wilson coefficients.

In summary, the SM basis contains ten operators in three sub-sectors, namely
\begin{align}
  \text{VLL},u \quad (\text{or VLL},c), \qquad
  \text{P}_\text{QCD}, \qquad
  \text{P}_\text{QED}.
\end{align}
Their chirality-flipped counterparts are absent in the SM basis because
no Wilson coefficients are generated for them due to the left-handed nature of
the SM electroweak interactions.

{\bf Beyond the SM}, it is convenient to introduce $d_i \to d_j \, d_i \bar d_i$
and $d_i \to d_j \, d_j \bar d_j$ operators with vector currents,
as the ones in Eq.~(4.2) of Ref.~\cite{Buras:2000if}
\begin{equation}
\begin{aligned}
  \OpL{11} = \; \OpL[\text{VLL},i+j]{1} &
  = (\bar d_j^\alpha \gamma_\mu P_L d_i^\alpha) \, \big[
    (\bar d_i^\beta \gamma^\mu P_L d_i^\beta)
  + (\bar d_j^\beta \gamma^\mu P_L d_j^\beta)\big] ,
\\
  \OpL{12} = \; \OpL[\text{VLR},i+j]{1} &
  = (\bar d_j^\alpha \gamma_\mu P_L d_i^\beta) \, \big[
    (\bar d_i^\beta \gamma^\mu P_R d_i^\alpha)
  + (\bar d_j^\beta \gamma^\mu P_R d_j^\alpha)\big] ,
\\
  \OpL{13} = \; \OpL[\text{VLR},i+j]{2} &
  = (\bar d_j^\alpha \gamma_\mu P_L d_i^\alpha) \, \big[
    (\bar d_i^\beta \gamma^\mu P_R d_i^\beta)
  + (\bar d_j^\beta \gamma^\mu P_R d_j^\beta)\big] ,
\end{aligned}
\end{equation}
and
\begin{equation}
\begin{aligned}
  \OpL{14} = \; \OpL[\text{VLL},i-j]{1} &
  = (\bar d_j^\alpha \gamma_\mu P_L d_i^\alpha) \,\big[
    (\bar d_i^\beta \gamma^\mu P_L d_i^\beta)
  - (\bar d_j^\beta \gamma^\mu P_L d_j^\beta)\big] ,
\\
  \OpL{15} = \; \OpL[\text{VLR},i-j]{1} &
  = (\bar d_j^\alpha \gamma_\mu P_L d_i^\beta) \, \big[
    (\bar d_i^\beta \gamma^\mu P_R d_i^\alpha)
  - (\bar d_j^\beta \gamma^\mu P_R d_j^\alpha)\big] ,
\\
  \OpL{16} = \; \OpL[\text{VLR},i-j]{2} &
  = (\bar d_j^\alpha \gamma_\mu P_L d_i^\alpha) \, \big[
    (\bar d_i^\beta \gamma^\mu P_R d_i^\beta)
  - (\bar d_j^\beta \gamma^\mu P_R d_j^\beta)\big] .
\end{aligned}
\end{equation}
Only the ``$i+j$'' operators mix into the penguin operators $\OpL{3,\ldots 10}$,
whereas the ``$i-j$'' ones do not. Analogous operators for $d_i \to d_j \, u_k \bar u_k$
are
\begin{equation}
\begin{aligned}
  \OpL{17} = \; \OpL[\text{VLR}, u-c]{1} &
  = (\bar d_j^\alpha \gamma_\mu P_L d_i^\beta) \, \big[
    (\bar u^\beta \gamma^\mu P_R \, u^\alpha)
  - (\bar c^\beta \gamma^\mu P_R \, c^\alpha) \big],
\\
  \OpL{18} = \; \OpL[\text{VLR}, u-c]{2} &
  = (\bar d_j^\alpha \gamma_\mu P_L d_i^\alpha) \, \big[
    (\bar u^\beta \gamma^\mu P_R \, u^\beta)
  - (\bar c^\beta \gamma^\mu P_R \, c^\beta) \big].
\end{aligned}
\end{equation}
They do not mix into the penguin operators $\OpL{3,\ldots 10}$ but
give threshold corrections when decoupling the charm quark.

The operators of the {\bf SRL sector} for $d_i \to d_j \, Q \oL{Q}$ with
$Q = u,c,d_{k\neq i, j}$ are given in Eqs.~(4.3) and (4.4) of Ref.~\cite{Buras:2000if}
\begin{align}
  \label{eq:def-BMU-SRL-ops}
  \OpL[\text{SRL}, Q]{1} &
  = (\bar d_j^\alpha P_R d_i^\beta) \, (\bar Q^\beta P_L \, Q^\alpha) ,
&
  \OpL[\text{SRL}, Q]{2} &
  = (\bar d_j^\alpha P_R d_i^\alpha) \, (\bar Q^\beta P_L \, Q^\beta) .
\end{align}
They are numbered as
\begin{equation}
\begin{aligned}
  (\OpL{19},\, \OpL{20}) & = (\OpL[\text{SRL}, u]{1},\, \OpL[\text{SRL}, u]{2}) ,
  \qquad \qquad
&
\\
  (\OpL{21},\, \OpL{22}) & = (\OpL[\text{SRL}, c]{1},\, \OpL[\text{SRL}, c]{2}) ,
  \qquad \qquad
&
  (\OpL{23},\, \OpL{24}) & = (\OpL[\text{SRL}, d_k]{1},\, \OpL[\text{SRL}, d_k]{2}) .
\end{aligned}
\end{equation}

The scalar operators of the {\bf SRR sectors} for $d_i \to d_j \, d_i \oL{d}_i$
and $d_i \to d_j \, d_j \oL{d}_j$ have been chosen in Ref.~\cite{Buras:2000if}
in such a way that they contain colour-singlet currents only, analogously
to the $\DF = 2$ case there. Due to the Fierz identities, there are only two
operators in the SRR,$i$ sub-sector
\begin{align}
  \label{eq:BMU-SRR,i}
  \OpL{25} = \OpL[\text{SRR}, i]{1} &
  = (\bar d_j^\alpha P_R \, d_i^\alpha) \, (\bar d_i^\beta P_R \, d_i^\beta) ,
&
  \OpL{26} = \OpL[\text{SRR}, i]{2} &
  = (\bar d_j^\alpha \sigma_{\mu\nu} P_R \, d_i^\alpha) \,
    (\bar d_i^\beta  \sigma^{\mu\nu} P_R \, d_i^\beta) ,
\intertext{and the SRR$,j$ one}
  \OpL{27} = \OpL[\text{SRR}, j]{1} &
  = (\bar d_j^\alpha P_R \, d_i^\alpha) \, (\bar d_j^\beta P_R \, d_j^\beta) ,
&
  \OpL{28} = \OpL[\text{SRR}, j]{2} &
  = (\bar d_j^\alpha \sigma_{\mu\nu} P_R \, d_i^\alpha) \,
    (\bar d_j^\beta  \sigma^{\mu\nu} P_R \, d_j^\beta) .
\end{align}

The case of $d_i \to d_j \, Q \oL{Q}$ with $Q = u,c,d_{k\neq i, j}$
comprises four operators per sub-sector
\begin{equation}
  \label{eq:def-BMU-SRR-ops}
\begin{aligned}
  \OpL[\text{SRR}, Q]{1} &
  = (\bar d_j^\alpha P_R \, d_i^\beta) \, (\bar Q^\beta P_R \, Q^\alpha) , \qquad
&
  \OpL[\text{SRR}, Q]{3} &
  = (\bar d_j^\alpha \sigma_{\mu\nu} P_R \, d_i^\beta) \,
    (\bar Q^\beta    \sigma^{\mu\nu} P_R \, Q^\alpha) ,
\\
  \OpL[\text{SRR}, Q]{2} &
  = (\bar d_j^\alpha P_R \, d_i^\alpha) \, (\bar Q^\beta P_R \, Q^\beta) ,
&
  \OpL[\text{SRR}, Q]{4} &
  = (\bar d_j^\alpha \sigma_{\mu\nu} P_R \, d_i^\alpha) \,
    (\bar Q^\beta    \sigma^{\mu\nu} P_R \, Q^\beta) .
\end{aligned}
\end{equation}
They are numbered consecutively
\begin{equation}
\begin{aligned}
  (\OpL{29},\, \OpL{30},\, \OpL{31},\, \OpL{32}) &
  = (\OpL[\text{SRR}, u]{1},\, \OpL[\text{SRR}, u]{2},\,
     \OpL[\text{SRR}, u]{3},\, \OpL[\text{SRR}, u]{4}) ,
\\
  (\OpL{33},\, \OpL{34},\, \OpL{35},\, \OpL{36}) &
  = (\OpL[\text{SRR}, c]{1},\,\, \OpL[\text{SRR}, c]{2},\,\,
     \OpL[\text{SRR}, c]{3},\,\, \OpL[\text{SRR}, c]{4}) ,
\\
  (\OpL{37},\, \OpL{38},\, \OpL{39},\, \OpL{40}) &
  = (\OpL[\text{SRR}, d_k]{1},\, \OpL[\text{SRR}, d_k]{2},\,
     \OpL[\text{SRR}, d_k]{3},\, \OpL[\text{SRR}, d_k]{4}) .
\end{aligned}
\end{equation}

Here, we use the standard definition
$\sigma_{\mu\nu} \equiv \frac{i}{2} [\gamma_\mu, \gamma_\nu]$,
contrary to Refs.~\cite{Buras:2000if,Ciuchini:1997bw} where
no overall factor of $i$ was present. It implies that our
BMU operators with tensor currents differ by a sign from
those in the original BMU paper~\cite{Buras:2000if}, which affects
a few signs in the ADMs of \refapp{app:ADM}.

As far as the chirality-flipped operators are concerned, their
numbering in the BMU basis is given by
\begin{align}
  \OpL{40 + i} & = \OpL{i}[P_L \leftrightarrow P_R] ,
\end{align}
i.e. they are found by interchanging $P_L \leftrightarrow P_R$ in the
``non-flipped'' operators.

%
%
%
\section{Anomalous dimensions}
\label{app:ADM}

In this appendix, we collect results of the LO and NLO QCD
ADMs. We set $\Nc = 3$ wherever the $N_c$-dependence is
provided in Ref.~\cite{Buras:2000if}, but keep the $\Nf$-dependence
explicit. Moreover, we provide the one-loop and
two-loop ADMs involving QED effects for those
sectors that mix into the QCD- and QED-penguin operators.

\vskip 0.2cm

The {\bf\bm VLL,$u$ sub-sector} (or, alternatively, VLL,$c$) contains two operators
$(\OpL[\text{VLL}, u]{1},\, \OpL[\text{VLL}, u]{2})$.
Their one- and two-loop QCD ADMs are given in Eqs.~(3.2, 3.3) of
Ref.~\cite{Buras:2000if}. For $\Nc=3$ and arbitrary~$\Nf$ they read
\begingroup
\setlength\arraycolsep{0.2cm}
\begin{align}
  \adm^{(10)}_{\text{VLL},u} &
  = \begin{pmatrix}
    -2 &  6  \\[0.4cm]
     6 & -2
  \end{pmatrix} ,
&
  \adm^{(20)}_{\text{VLL},u} &
  = \begin{pmatrix}
    -\frac{21}{2} - \frac{2}{9} \Nf &   \frac{7}{2} + \frac{2}{3} \Nf \\[0.4cm]
     \frac{7}{2}  + \frac{2}{3} \Nf & -\frac{21}{2} - \frac{2}{9} \Nf
  \end{pmatrix} .
\end{align}
\endgroup
Since the VLL,$u$ operators are part of the SM operator basis, their QED ADMs
are known, too~\cite{Shifman:1976ge, Gilman:1979bc, Buras:1992zv, Ciuchini:1993vr}
\begingroup
\setlength\arraycolsep{0.2cm}
\begin{align}
  \adm^{(01)}_{\text{VLL},u} &
  = \begin{pmatrix}
    -\frac{8}{3} & 0  \\[0.4cm]
    0 & -\frac{8}{3}
  \end{pmatrix} ,
&
  \adm^{(11)}_{\text{VLL},u} &
  = \begin{pmatrix}
     \frac{194}{9} & -\frac{2}{3} \\[0.4cm]
     \frac{25}{3}  & -\frac{49}{9}
  \end{pmatrix} .
\end{align}
\endgroup

\vskip 0.2cm

The {\bf\bm sub-sectors VLR,$i\pm j$ and VLR,$u-c$} contain two operators
$(\OpL[\text{VLR}, Q_i \pm Q_j]{1},\, \OpL[\text{VLR}, Q_i \pm Q_j]{2})$.
Their one- and two-loop QCD ADMs can be derived directly from the VLR sector
given in Eqs.~(3.4, 3.5) of Ref.~\cite{Buras:2000if}, see also the explanation around
Eq.~(4.7) in Ref.~\cite{Buras:2000if}. For $N_c = 3$ and arbitrary $\Nf$ one finds
\begingroup
\setlength\arraycolsep{0.2cm}
\begin{align}
  \adm^{(10)}_{\text{VLR},Q_i\pm Q_j} &
  = \begin{pmatrix}
  -16 & 0  \\[0.4cm]
  -6  & 2
  \end{pmatrix} ,
&
  \adm^{(20)}_{\text{VLR},Q_i\pm Q_j} &
  = \begin{pmatrix}
    -\frac{1343}{6} + \frac{68}{9} \Nf & -\frac{225}{2} + 4 \Nf           \\[0.4cm]
    -99 + \frac{22}{3}  \Nf            &  \frac{71}{3}  - \frac{22}{9} \Nf
  \end{pmatrix} ,
\end{align}
\endgroup
for $Q_i \pm Q_j = \{ d_i\pm d_j,\; u-c \}$. For completeness, we provide
here also the LO and NLO ADMs involving QED effects for $Q_i + Q_j$. They read
\begingroup
\setlength\arraycolsep{0.2cm}
\begin{align}
  \adm^{(01)}_{\text{VLR},Q_i+ Q_j} &
  = \begin{pmatrix}
    -\frac{4}{3} & 0 \\[0.4cm]
    0 & -\frac{4}{3}
  \end{pmatrix} ,
&
  \adm^{(11)}_{\text{VLR},Q_i+ Q_j} &
  = \begin{pmatrix}
    -\frac{2}{3} \Nc - \frac{10}{3 \Nc} & 4 \\[0.4cm]
    -\frac{2}{3} & 4 \Nc - \frac{10}{3 \Nc}
  \end{pmatrix} .
\end{align}
\endgroup
They have been derived from Ref.~\cite{Buras:1992zv} using the fact that
only the current-current insertions are needed, and considering the special case
$\Nd = 2$ and $\Nu = 0$.

\vskip 0.2cm

The {\bf\bm sub-sectors VLL,$i\pm j$} mediate $d_i \to d_j\, d_i\bar d_i$ and
$d_i \to d_j\, d_j\bar d_j$ that both contain a single operator
$\OpL[\text{VLL},i+j]{1}$ and $\OpL[\text{VLL},i-j]{1}$, respectively.
The corresponding LO and NLO ADMs in QCD are given by the $\DF=2$ result
in Eq.~(2.21) of Ref.~\cite{Buras:2000if}
\begin{align}
  \adm^{(10)}_{\text{VLL},i\pm j} & = \left( 4 \right),
&
  \adm^{(20)}_{\text{VLL},i\pm j} &
  =  \left( -7 + \frac{4}{9} \Nf \right) .
\end{align}

For completeness, we provide here also the LO and NLO ADMs involving
QED effects for $\OpL[\text{VLL},i+j]{1}$. They read
\begin{align}
  \adm^{(01)}_{\text{VLL},i+j} &
  = \left( \frac{4}{3} \right) ,
&
  \adm^{(11)}_{\text{VLL},i+j} &
  = \left( \frac{38}{3\Nc} - 4\Nc - \frac{26}{3} \right) .
\end{align}

\vskip 0.2cm

The $8\times 8$ ADMs $\adm^{(mn)}_P$ of {\bf QCD- and QED-penguin operators}
is known~\cite{Buras:1991jm, Buras:1992tc, Buras:1992zv, Ciuchini:1993vr} for
arbitrary numbers of active flavours $\Nf=\Nu+\Nd$ at one- and two-loops, for all the contributions
$mn = 10, 20, 01, 11$ in Eq.~\eqref{eq:adm-exp}.\footnote{The three-loop NNLO QCD
result $\adm^{(30)}$ for the mixing of the VLL,$u$ sub-sector and the QCD-penguin
operators is known from Ref.~\cite{Gorbahn:2004my}.} Similarly, the mixing
of current-current operators VLL,$u$ into the penguin ones
is known from the SM calculations, too. We therefore refrain from
repeating these results here.
The mixing of the operators $(Q_{11}, Q_{12}, Q_{13})$ from the sub-sectors VLL,$i+j$
and VLR,$i+j$ into the penguin operators $(Q_3, \ldots, Q_{10}$) has
been calculated in Ref.~\cite{Buras:2000if} up to the NLO in QCD, giving
\begingroup
\setlength\arraycolsep{0.2cm}
\begin{align}
  \adm^{(10)}_{i+j \to \text{P}} &
  = \frac{4}{3} \begin{pmatrix}
    -\frac{1}{\Nc} & 1 & -\frac{1}{\Nc} & 1 & 0 & 0 & 0 & 0 \\[0.3cm]
    -\frac{1}{\Nc} & 1 & -\frac{1}{\Nc} & 1 & 0 & 0 & 0 & 0 \\[0.3cm]
    0  & 0 & 0 & 0 & 0 & 0 & 0 & 0
  \end{pmatrix} ,
\\[0.3cm]
  \adm^{(20)}_{i+j \to \text{P}} &
  = \begin{pmatrix}
    \frac{3538}{243} & \frac{2654}{81} & -\frac{6218}{243} & \frac{1754}{81} &
    0 & 0 & 0 & 0 \\[0.3cm]
   -\frac{1364}{243} & \frac{212}{81}  &  \frac{1408}{243} & \frac{1472}{81} &
    0 & 0 & 0 & 0 \\[0.3cm]
   -\frac{122}{9}    & -\frac{22}{3}   & \frac{166}{9}     & -\frac{22}{3}   &
    0 & 0 & 0 & 0
  \end{pmatrix} .
\end{align}
\endgroup
However, our results below for their LO QED and NLO QED$\times$QCD ADMs are new
\begingroup
\setlength\arraycolsep{0.2cm}
\begin{align}
  \adm^{(01)}_{i+j \to \text{P}} &
  = - \frac{16}{27} \begin{pmatrix}
    0 & 0 & 0 & 0 & 1 + \Nc & 0 & 1 + \Nc & 0 \\[0.3cm]
    0 & 0 & 0 & 0 &       1 & 0 &       1 & 0 \\[0.3cm]
    0 & 0 & 0 & 0 &     \Nc & 0 &     \Nc & 0
  \end{pmatrix} ,
\\[0.3cm]
  \adm^{(11)}_{i+j \to \text{P}} &
  = -\frac{8}{27} \begin{pmatrix}
    -\frac{29}{9 \Nc}-\frac{17}{9} & -\frac{53}{9 \Nc} & -\frac{17}{9}
    \\[0.2cm]
    \frac{17 \Nc}{9}+\frac{29}{9} & \frac{53}{9} & \frac{17 \Nc}{9}
    \\[0.2cm]
    -\frac{29}{9 \Nc}-\frac{17}{9} & -\frac{53}{9 \Nc} & -\frac{17}{9}
    \\[0.2cm]
    \frac{17 \Nc}{9}+\frac{29}{9} & \frac{53}{9} & \frac{17 \Nc}{9}
    \\[0.2cm]
    3 \Nc^2-\frac{58 \Nc}{9}+\frac{130}{9 \Nc}-14 & -\frac{106 \Nc}{9}-\frac{254}{9 \Nc} & 3 \Nc^2-14
    \\[0.2cm]
    11 \Nc-8 & 40 & 11 \Nc
    \\[0.2cm]
    3 \Nc^2-\frac{58 \Nc}{9}-\frac{230}{9 \Nc}-4  & \frac{250}{9 \Nc}-\frac{106 \Nc}{9} & 3 \Nc^2-4
    \\[0.2cm]
    \Nc+32 & -16 & \Nc
  \end{pmatrix}^{T}
  \label{eq:gam11_ij-P}
\\[0.3cm] &
  = -\frac{8}{27} \begin{pmatrix}
    -\frac{80}{27} & \frac{80}{9} & -\frac{80}{27}  & \frac{80}{9} &
    -\frac{41}{27} & 25           & -\frac{131}{27} & 35   \\[0.3cm]
    -\frac{53}{27} & \frac{53}{9} & -\frac{53}{27}  & \frac{53}{9} &
    -\frac{1208}{27} & 40         & -\frac{704}{27} & -16 \\[0.3cm]
    -\frac{17}{9}  & \frac{17}{3} & -\frac{17}{9}   & \frac{17}{3} &
     13            & 33           & 23              & 3
  \end{pmatrix} .
\end{align}
\endgroup
Note the transposition of the matrix on the r.h.s.\ of Eq.~\eqref{eq:gam11_ij-P}.
They have been derived from Ref.~\cite{Buras:1992zv}, subtracting the contributions
due to current-current insertions from the full results for the special case
$\Nd = 2$ and $\Nu =0$ to retain only the mixing due to penguin insertions.

\vskip 0.2cm

The {\bf\bm SRR,$i(j)$ sub-sectors} contain only two operators
$(\OpL[\text{SRR}, i(j)]{1},\, \OpL[\text{SRR}, i(j)]{2})$, in contrast to
four operators in SRR,$Q$, because two of them can be eliminated using the Fierz relations.
The ADMs of the SRR,$i(j)$ sub-sector for $\DF=1$ are the same as
for $\DF = 2$ in Eqs.~(2.18) and (2.20) of Ref.~\cite{Buras:2000if}
\begingroup
\setlength\arraycolsep{0.2cm}
\begin{align}
  \label{eq:adm_SRRi}
  \adm^{(10)}_{\text{SRR},i(j)} &
  = \begin{pmatrix}
    -10 & -\frac{1}{6} \\[0.4cm]
     40 &  \frac{34}{3}
  \end{pmatrix} ,
&
  \adm^{(20)}_{\text{SRR},i(j)} &
  = \begin{pmatrix}
   -\frac{1459}{9} + \frac{74}{9}  \Nf & \frac{35}{36}  + \frac{1}{54} \Nf \\[0.3cm]
    \frac{6332}{9} - \frac{584}{9} \Nf & \frac{2065}{9} - \frac{394}{27} \Nf
  \end{pmatrix} ,
\end{align}
\endgroup
up to overall signs in the off-diagonal entries, which is due to the $\sigma_{\mu\nu}$ normalization issue,
as described at the end of \refapp{app:non-leptonic-op's}.

\vskip 0.2cm

The {\bf\bm SRR,$Q$ sub-sectors} with $Q = u, c, d_{k\neq i, j}$ contain four
operators $(\OpL[\text{SRR}, Q]{1},\, \ldots,\, \OpL[\text{SRR}, Q]{4})$.
Their one- and two-loop QCD ADMs are given in Eqs.~(3.8, 3.9) of Ref.~\cite{Buras:2000if}.
For $\Nc=3$ and arbitrary~$\Nf$ they read
\begingroup
\setlength\arraycolsep{0.18cm}
\begin{align}
  \label{eq:adm_SRR_00}
  \adm^{(10)}_{\text{SRR},Q} &
  = \begin{pmatrix}
    2   &  -6 &  -\frac{7}{6} & -\frac{1}{2} \\[0.3cm]
    0   & -16 &            -1 &  \frac{1}{3} \\[0.3cm]
    -56 & -24 & -\frac{38}{3} &            6 \\[0.3cm]
    -48 &  16 &             0 & \frac{16}{3}
  \end{pmatrix} ,
\\[0.2cm]
  \label{eq:adm_SRR_10}
  \adm^{(20)}_{\text{SRR},Q} & = \begin{pmatrix}
  \frac{350}{9}-\frac{64}{9} \Nf     & -\frac{470}{3}+\frac{16}{3} \Nf   &
 -\frac{805}{36}+\frac{7}{54} \Nf    &  \frac{77}{12}+\frac{1}{18} \Nf   \\[0.4cm]
 -\frac{130}{3}                      & -\frac{2710}{9}+\frac{80}{9} \Nf  &
 -\frac{31}{2}+\frac{1}{9} \Nf       &  \frac{61}{18}-\frac{1}{27} \Nf   \\[0.4cm]
 -\frac{12292}{9}+\frac{616}{9} \Nf  & -\frac{2908}{3}+\frac{88}{3} \Nf  &
 -\frac{1262}{9}+\frac{200}{27} \Nf  &  50-\frac{8}{3} \Nf               \\[0.4cm]
 -\frac{1880}{3}+\frac{176}{3} \Nf   &  \frac{2648}{9}-\frac{176}{9} \Nf &
  \frac{26}{3}+\frac{8}{3} \Nf       &  \frac{1582}{9}-\frac{232}{27} \Nf
  \end{pmatrix} .
\end{align}
\endgroup
Again, overall sign differences w.r.t.\ to Ref.~\cite{Buras:2000if} show up in the
off-diagonal $2\times2$ blocks due to the $\sigma_{\mu\nu}$ normalization issue.

\vskip 0.2cm

The {\bf\bm SRL,$Q$ sub-sectors} with $Q = u, c, d_{k\neq i, j}$
contain two operators $(\OpL[\text{SRL}, Q]{1},\quad \OpL[\text{SRL}, Q]{2})$.
Their one- and two-loop QCD ADMs are given in Eqs.~(3.6, 3.7) of Ref.~\cite{Buras:2000if}.
For $\Nc=3$ and arbitrary~$\Nf$ they read
\begingroup
\setlength\arraycolsep{0.2cm}
\begin{align}
  \adm^{(10)}_{\text{SRL},Q} &
  = \begin{pmatrix}
    2 & -6  \\[0.4cm]
    0 & -16
  \end{pmatrix}\,,
&
  \adm^{(20)}_{\text{SRL},Q} &
  = \begin{pmatrix}
  \frac{71}{3} - \frac{22}{9} \Nf & -99 + \frac{22}{3} \Nf            \\[0.4cm]
  -\frac{225}{2} + 4 \Nf          & -\frac{1343}{6} + \frac{68}{9} \Nf
  \end{pmatrix}\,.
\end{align}
\endgroup

%
%
%
\section{Threshold corrections}
\label{app:threshold}

In \refsec{sec:RG}, threshold corrections that arise when going from the $\Nf$- to the
$(\Nf-1)$-flavour WET have been discussed in general terms. They are described
by the matrices $\hat M^{(\Nf)}(\mu_f)$ defined in Eq.~\eqref{eq:def-thr-corr}.
These matrices are obtained by matching matrix elements of operators
of the $N_f$- and the $(N_f-1)$-flavour WET, which is marked by the
superscript $\Nf$. They are not quadratic matrices because the number of
linearly independent operators becomes smaller when one of the quarks gets
decoupled, as discussed at the end of \refsec{sec:RG}. Their perturbative expansion reads
\begin{equation}
\begin{aligned}
  \hat M^{(\Nf)}(\mu_\Nf) &
  = \hat M_{00}^{(\Nf)}
  + \frac{\alS(\mu_\Nf)}{4\pi} \hat M_{10}^{(\Nf)} (\mu_\Nf)
  + \frac{\alE(\mu_\Nf)}{4\pi} \hat M_{01}^{(\Nf)} (\mu_\Nf)
  + \ldots ,
\end{aligned}
\end{equation}
where the first term arises at the tree-level,
while the remaining two are due to one-loop QCD and QED corrections, respectively.

For the SM-generated operators $Q_1$-$Q_{10}$ and their chirality-flipped counterparts,
the $10\times 10$ tree-level matching matrix blocks are equal to unit matrices,\footnote{
Here and below we shall use operator names as indices of the matching matrices,
to avoid confusion in the $\Nf =3,4$ cases.}
i.e. $[\hat M_{00}^{(\Nf)}]_{\OpL{i},\,\OpL{j}} = \delta_{ij}$.
This is one of the advantages of using a redundant operator set in the SM-like sector of the BMU basis
for $\Nf =3,4$. However, as far as the BSM operators are concerned, we shall remove all the linearly dependent ones.
In particular, Eqs.~\eqref{eq:nf4-2} and~\eqref{eq:nf4-3} give rise to
the following non-vanishing entries in $[\hat M_{00}^{(5)}]$:
\begin{align}
    [\hat M_{00}^{(5)}]_{\OpL{3},\,\OpL{11}} &
  = [\hat M_{00}^{(5)}]_{\OpL{5},\,\OpL{13}}
  = [\hat M_{00}^{(5)}]_{\OpL{6},\,\OpL{12}} = + \frac{2}{3} ,
\\
    [\hat M_{00}^{(5)}]_{\OpL{7},\,\OpL{13}} &
  = [\hat M_{00}^{(5)}]_{\OpL{8},\,\OpL{12}}
  = [\hat M_{00}^{(5)}]_{\OpL{9},\,\OpL{11}} = - \frac{2}{3} .
\end{align}
In the $[\hat M_{00}^{(4)}]$ case, tree-level corrections due to
Eq.~\eqref{eq:nf3-3} imply that
\begin{align}
    [\hat M_{00}^{(4)}]_{\OpL{5},\,\OpL{18}} &
  = [\hat M_{00}^{(4)}]_{\OpL{6},\,\OpL{17}} = \frac{1}{3} ,
&
    [\hat M_{00}^{(4)}]_{\OpL{7},\,\OpL{18}} &
  = [\hat M_{00}^{(4)}]_{\OpL{8},\,\OpL{17}} = \frac{2}{3} .
\end{align}
Analogous relations hold in the chirality-flipped sector. They are obtained
by simply replacing $\OpL{i} \to \OpL{40 + i}$.

Beyond the tree level, the $b$- and $c$-quark decoupling proceeds
through insertions of operators that contain pairs of these quarks into
penguin diagrams, as shown in \reffig{fig:peng-ins-1l}. In the VLL and VLR sectors,
they arise mainly from the QCD- and QED-penguin operators. They are thus known
from the ancient SM calculation~\cite{Buras:1993dy}.
The $10\times 10$ matrices $\hat M_{10}^{(5)}$ and $\hat M_{10}^{(4)}$ for
the operators $i \in \{\OpL{1},\ldots,\OpL{10} \}$ have the following
non-vanishing columns:
\begin{align}
  \label{eq:a.1}
  \mu_5: & &
    [\hat M_{10}^{(5)}]_{i, \OpL{4}} = [\hat M_{10}^{(5)}]_{i, \OpL{6}}
  = -2 \, [\hat M_{10}^{(5)}]_{i, \OpL{8}}
  = -2 \, [\hat M_{10}^{(5)}]_{i, \OpL{10}} &
  = \left(-\frac{5}{9} + \frac{2}{3} \ln \frac{m_b}{\mu_5} \right) \, P_i \,,
\\[0.2cm]
  \label{eq:a.2}
  \mu_4: & &
    [\hat M_{10}^{(4)}]_{i, \OpL{4}} = [\hat M_{10}^{(4)}]_{i, \OpL{6}}
  = [\hat M_{10}^{(4)}]_{i, \OpL{8}} = [\hat M_{10}^{(4)}]_{i, \OpL{10}} &
  = \left(-\frac{5}{9} + \frac{2}{3} \ln \frac{m_c}{\mu_4} \right) \, P_i \,,
\end{align}
where
\begin{equation}
  \label{eq:a.3}
  P_i = \big( \,
  0,\; 0,\; -\frac{1}{\Nc},\; 1,\; -\frac{1}{\Nc},\; 1,\; 0,\; 0,\; 0,\; 0 \,
  \big)_i \, .
\end{equation}

The corresponding results for the QED matrices $\hat M_{01}^{(5)}$ and
$\hat M_{01}^{(4)}$ are given as~\cite{Buras:1993dy}
\begin{align}
  \mu_5: & &
  \frac{1}{2\Nc} [\hat M_{01}^{(5)}]_{i, \OpL{3}}
  = \frac{1}{2} [\hat M_{01}^{(5)}]_{i, \OpL{4}}
  = \frac{1}{2\Nc} [\hat M_{01}^{(5)}]_{i, \OpL{5}}
  = \frac{1}{2} [\hat M_{01}^{(5)}]_{i, \OpL{6}} & =
\notag
\\
  & &
    -\frac{1}{\Nc} [\hat M_{01}^{(5)}]_{i, \OpL{7}}
  = -    [\hat M_{01}^{(5)}]_{i, \OpL{8}}
  = -\frac{1}{\Nc} [\hat M_{01}^{(5)}]_{i, \OpL{9}}
  = -    [\hat M_{01}^{(5)}]_{i, \OpL{10}} &
  = \left(\frac{10}{81} - \frac{4}{27} \ln \frac{m_b}{\mu_5} \right) \, \oL{P}_i \,,
\\[0.2cm]
  \mu_4: & &
    \frac{1}{\Nc} [\hat M_{01}^{(4)}]_{i, \OpL{3}} = [\hat M_{01}^{(4)}]_{i, \OpL{4}}
  = \frac{1}{\Nc} [\hat M_{01}^{(4)}]_{i, \OpL{5}} = [\hat M_{01}^{(4)}]_{i, \OpL{6}} & =
\notag
\\
  & &
    \frac{1}{\Nc} [\hat M_{01}^{(4)}]_{i, \OpL{7}} = [\hat M_{01}^{(4)}]_{i, \OpL{8}}
  = \frac{1}{\Nc} [\hat M_{01}^{(4)}]_{i, \OpL{9}} = [\hat M_{01}^{(4)}]_{i, \OpL{10}} &
  = \left(-\frac{40}{81} + \frac{16}{27} \ln \frac{m_c}{\mu_4} \right) \, \oL{P}_i \,,
\end{align}
where
\begin{equation}
  \label{eq:a.3-QED}
  \oL{P}_i = \big( \,
  0,\; 0,\; 0,\; 0,\; 0,\; 0,\; 1,\; 0,\; 1,\; 0 \,
  \big)_i \, .
\end{equation}

Among the BSM operators in the BMU basis, only the VLR,$u-c$
operator $\OpL{17}$ gives one-loop threshold corrections
when decoupling the charm quark. They are given by
\begin{align}
  \label{eq:a.4}
  \mu_4: & &
  [\hat M_{10}^{(4)}]_{i, \OpL{17}} &
  = \left(\frac{5}{9} - \frac{2}{3} \ln \frac{m_c}{\mu_4} \right) \, P_i \, ,
\end{align}
\begin{align}\label{eq:a.5}
  \mu_4: & &
  [\hat M_{01}^{(4)}]_{i, \OpL{17}} = \frac{1}{\Nc} [\hat M_{01}^{(4)}]_{i, \OpL{18}}
  = \left(\frac{40}{81} - \frac{16}{27} \ln \frac{m_c}{\mu_4} \right) \, \oL{P}_i \, ,
\end{align}
in the QCD and QED cases, respectively. One should remember though
that $\OpL{17}$ above the charm threshold does not mix into the penguin
operators. Consequently, its effect on Kaon physics originates solely from the
one-loop matching corrections \eqref{eq:a.4} and \eqref{eq:a.5},
as long as the physical amplitudes are calculated in the $\Nf=3$ WET framework.

The SRR,$b$ operators with $\sigma_{\mu\nu}$ give rise to threshold corrections
to the chromo-magnetic dipole operators only, see for example
Refs.~\cite{Borzumati:1999qt, Bobeth:2011st}.
Other SRR,$b$ and SRL,$b$ operators will not generate threshold corrections.
The same comment applies to the corresponding SRR,$c$ and SRL,$c$ operators.

%
%
%
%
%
%
%

%
%

\newpage
\renewcommand{\refname}{R\lowercase{eferences}}

\addcontentsline{toc}{section}{References}

\bibliographystyle{JHEP}

\small

\bibliography{Bookallrefs}

\end{document}